\documentclass[manuscript, screen, nonacm]{acmart}

\usepackage{natbib}
\usepackage{fancyvrb}
\usepackage{fvextra}
\usepackage{xcolor}
\usepackage{tcolorbox}
\usepackage{tabularx}
\usepackage{array}
\usepackage{colortbl}
\usepackage{pdflscape}
\usepackage{array}
\usepackage{graphicx}
\usepackage{longtable}
\usepackage{geometry}
\usepackage[pro]{fontawesome5}

\setcopyright{none}
\renewcommand\footnotetextcopyrightpermission[1]{} %

\usepackage{verbatim}

\usepackage{xspace}
\xspaceaddexceptions{]\}}

\usepackage{xcolor}
\usepackage{subcaption}
\usepackage{geometry}
\usepackage {enumitem}

\definecolor{darkgreen}{rgb}{0.05,0.5,0.05}

\newcommand{\Vanilla}{Vanilla\xspace}

\newcommand{\CMS}{GEMS\xspace}

\definecolor{waterblue}{RGB}{173,216,230}
\definecolor{grassgreen}{RGB}{231,255,201}

\newcommand{\CMSIcon}{{\setlength{\fboxsep}{1.5pt}\fcolorbox{gray}{waterblue}{\textcolor{black}{\(\mathsf{G}\)}}}\xspace}
\newcommand{\vanillaIcon}{{\setlength{\fboxsep}{1.5pt}\fcolorbox{gray}{grassgreen}{\textcolor{black}{\(\mathsf{V}\)}}}\xspace}
\newcommand{\codeIcon}{{\tiny\faCode}\hspace{0.25em}}

\definecolor{grayline}{gray}{0.5}

\begin{document}

\title{GEMS: Generative Expert Metric System through Iterative Prompt Priming}

\author{Ti-Chung Cheng}
\affiliation{%
  \institution{University of Illinois Urbana-Champaign}
  \city{Urbana-Champaign}
  \country{USA}
  }
\email{tcheng10@illinois.edu}

\author{Carmen Badea}
\affiliation{%
  \institution{Microsoft Research}
  \city{Redmond}
  \country{USA}
  }
\email{cabadea@microsoft.com}

\author{Christian Bird}
\affiliation{%
  \institution{Microsoft Research}
  \city{Redmond}
  \country{USA}
  }
\email{cbird@microsoft.com}

\author{Thomas Zimmermann}
\affiliation{%
  \institution{Microsoft Research}
  \city{Redmond}
  \country{USA}
  }
  \email{tzimmer@microsoft.com}
\email{cbird@microsoft.com}

\author{Robert DeLine}
\affiliation{%
  \institution{Microsoft Research}
  \city{Redmond}
  \country{USA}
  }
\email{rdeline@microsoft.com}

\author{Nicole Forsgren}
\affiliation{%
  \institution{Microsoft Research}
  \city{Redmond}
  \country{USA}
  }
\email{niforsgr@microsoft.com}

\author{Denae Ford}
\affiliation{%
  \institution{Microsoft Research}
  \city{Redmond}
  \country{USA}
  }
\email{denae@microsoft.com}

\renewcommand{\shortauthors}{Cheng et al.}

\begin{abstract}
  Across domains, metrics and measurements are fundamental to identifying challenges, informing decisions, and resolving conflicts. Despite the abundance of data available in this information age, not only can it be challenging for a single expert to work across multi-disciplinary data~\cite{maoHowDataScientistsWork2019}, but non-experts can also find it unintuitive to create effective measures or transform theories into context-specific metrics that are chosen appropriately~\cite{grayMeasurementMadnessRecognizing2014}. This technical report addresses this challenge by examining software communities within large software corporations, where different measures are used as proxies to locate counterparts within the organization to transfer tacit knowledge. We propose a prompt-engineering framework inspired by neural activities, demonstrating that generative models can extract and summarize theories and perform basic reasoning, thereby transforming concepts into context-aware metrics to support software communities given software repository data. While this research zoomed in on software communities, we believe the framework's applicability extends across various fields, showcasing expert-theory-inspired metrics that aid in triaging complex challenges.
\end{abstract}

\maketitle

\section{Introduction}
Across the industry, many engineering teams express a desire to understand and improve their software practices~\cite{raybournDatadrivenApproachRethinking2021}. However, consistently and accurately measuring software engineering has remained elusive~\cite{forsgrenSPACEDeveloperProductivity2021}. Measuring engineering's processes, artifacts, collaborations, and even impact can be challenging. Yet attempts to create measures of software engineering have always been around: most computer science students have seen the picture of Margaret Hamilton next to a stack of paper that stands taller than her, containing the printed code she wrote for the Apollo program. While these stacks of paper do not represent the impact of her work, they are some symbols of her work (the authors maintain that lines of code, or reams of code printouts, are not good metrics for measuring coding output).

Many engineering teams and organizations struggle to understand and improve their software practices. When faced with questions about impact or collaboration, what data should be used? When faced with questions about software, many engineers resort to data that is easily available and instrumented in their systems, for example, pull requests and commits (e.g., \cite{VasilescuQualityProductivity2015}). While these data can be useful in many cases, they are not always suitable proxies. For example, the number of pull requests can miss the complexity or difficulty of the work done; in another example, the best solution may actually delete code, which would reflect "negative" productivity if simply counting lines of code. In our experience, we have seen practitioners make these mistakes quite often, reaching for available data rather than taking a thoughtful approach to operationalizing the concept they are trying to measure and test. This is understandable! They are trained as programmers and engineers, not as researchers. But the impacts of these mistakes can be significant for organizations and developers: For example, if work effort is measured incorrectly (whether coding, code reviews, debugging, or other invisible work such as unblocking others' work), it can lead to poor resource allocation, bad work design, burnout, and attrition.

In the face of limited experience designing and conducting analyses and, furthermore, limited data availability, many engineering teams still want to improve. One common way to do this is by learning from others~\cite{laveSituatingLearningCommunities1991}. However, this can still pose challenges because knowledge transfer is often most useful when done between similar contexts~\cite{laveSituatingLearningCommunities1991}; in engineering, this could be work processes, technology stacks, or programming languages used. 

To accomplish this, individuals and teams need to locate the right person to pose the right questions to, often beginning by relying on local networks as early computer-supported cooperative work research by~\citet{mcdonaldJustTalkMe1998} illuminated the complex processes of expertise identification and selection when locating experts to accomplish a task. They underscored the necessity for individuals to identify the skills and expertise of others before consulting the most suitable expert. In software developer communities this challenge is further compounded by the scale and complexity of software systems and large-scale collaboration. Software engineers and technical leads often have ambiguous queries or objectives that require specific skill sets and search heuristics to locate key information to achieve these goals. For instance, questions like: ``How should I improve my team's new hire experiences?'' necessitate a nuanced understanding of developer operations, organizational management, processes, and communication, thereby challenging the refinement of questions and the narrowing of search criteria, especially when incorporating non-technical search parameters~\cite{kintabRecommendingSoftwareExperts}.

Subsequent studies, like those by~\citet{yaroshAskingRightPerson2012}, investigated how to assist individuals in locating these experts. However, many challenges faced by individuals cannot be addressed by a single expert. Instead, solutions often require the collective expertise and guidance of small communities or project-based teams. For example, a security developer might want to learn, ``How can I design test cases that are ethical and just?'' which would benefit from engaging with a panel of colleagues experienced in both ethics and security protocols.

Moreover, finding counterparts inside a company or developer community that understand the context of a given question is fraught with difficulty. Prior research, such as that by~\citet{elliottFreeSoftwareDevelopers2003}, highlights the emergence of subcultures and common jargon among developers, leading to the formation of new sub-communities as organizations expand. This phenomenon poses a challenge for semantic matching, as the phrases and contexts embedded in questions require a nuanced and deeper understanding. In addition, as organizations and developer communities grow and change, an individual's personal network is unlikely to keep up with the new growth, creating inherent limitations to "simply" asking someone for help. That is, even if they know the right questions to ask, they may not know the people exist, or the people they would have asked may have left the organization or community~\cite{latozaHardtoanswerQuestionsCode2010}.

With these challenges in mind, we explored the following research question: \textbf{How can we leverage Large Foundation Models (LFMs) to match teams within software corporations to improve developer team performance?}

To answer this question, we explored and developed a prototype targeting software communities to automate expertise identification and selection within these software communities. We built this prototype using 
open-source software communities in order to support replication and eliminate privacy concerns. The results from our prototype identify metrics used for expert identification, thus providing greater visibility into the process, and are likely to transfer across domains (e.g. teams within software corporations), making it highly generalizable.

This technical report begins in Section~\ref{sec:prob-def} by defining the problem statement. Section~\ref{sec:sys-design} elaborates on the technical design and architecture of this generative-based system. We present initial outcomes and case examples in Section~\ref{sec:findings} and discuss the strengths, weaknesses, and opportunities of developing this prototype in Section~\ref{sec:discussion}. 

This technical report discusses two major contributions: (1) this work demonstrates it is possible to leverage LFMs to create generalized and scalable systems to identify expertise with a particular goal in mind (in our explorations, improving developer productivity), and (2) we highlight an architectural approach and several key techniques that enable an LFM-powered system to provide usable results grounded in expert literature. We believe that incorporating LFMs is an innovative approach to addressing the challenges of expertise identification and community matching, and shows promise in helping individuals and teams in OSS communities and corporate networks identify relevant expertise across weak ties.

\section{Problem Definition}\label{sec:prob-def}
We begin by splitting the research question: \textbf{"How can we leverage Large Foundation Models (LFMs) to match teams within software corporations to improve developer team performance?"} into three key components: (a) enhancing developer team performance, (b) team matching, and (c) the application of LFMs.

To address the improvement of developer team performance, we referenced prior literature covering decades of research which highlights challenges such as individual information needs during software engineering tasks~\cite{koInformationNeedsCollocated2007}, communication issues within and between software teams, and the complexities of navigating software systems~\cite{poileCoordinationLargeScaleSoftware}. Furthermore, efforts to define holistic measures for assessing developer productivity are noted~\cite{forsgrenSPACEDeveloperProductivity2021}. Despite the valuable directions provided by these studies, the development of tailored solutions for specific team challenges remains prohibitively expensive. We define a high-level software engineering challenge as $C$, which may be so ambiguous due to the broad nature of the challenge that an individual would not be able to come up with solutions without in-depth research into its specifics. We introduce a goal metric ($M_g$) as a proxy for optimizing the resolution or mitigation of $C$. For the purpose of our system prototype, we limit $C$ to a single $M_g$, while acknowledging that addressing $C$ fully may require improving multiple goal metrics.

Understanding that numerous factors might influence $M_g$, we identify a set of supporting metrics ($M_s$) believed to impact $M_g$ indirectly or directly. It's crucial to recognize $M_g$ and $M_s$ not as definitive metrics, but as proxies for better measurement and quantification of the targeted metrics.

In our efforts to improve developer team productivity, we adopt the communities of practice model advocated by~\citet{laveSituatingLearningCommunities1991}, which fosters regular interaction among individuals with a common interest or concern. This approach proves beneficial, particularly when addressing challenges that stem from institutional or tacit knowledge, which are not easily conveyed through documentation, or widely understood by general experts. We propose not just a match but a partnership between two teams: $T_x$, requiring support to address $C$, and  $T_y$, a team that is deemed capable of providing support and improvement of $M_g$ for team $T_x$.

Furthermore, we aim to harness LFMs for synthesizing and utilizing a broad spectrum of text-based information to develop metrics covering $M_g$ and the set of $M_s$ that helps identify $T_x$ and $T_y$. This text-based information includes but is not limited to: code artifacts, system documentation, markdown, git commits, and measures of interactivity. With the advancements in LFMs~\cite{bubeckSparksArtificialGeneral2023}, we leverage their contextual understanding, built-in logical reasoning capabilities, and ability to develop code. The LFM-based prototype system we developed is denoted as Generative Expert Metric System (GEMS).

With these definitions established, given available data ($D$), we articulate that the prototype demonstrates the system's ability to facilitate the following pipeline: 
$GEMS(D, C) \rightarrow (M_g, T_x) \rightarrow ((M_{s1}, M_{s2}, \ldots, M_{sn}), T_y)$.

\section{Approach and System Design}\label{sec:sys-design}
In this technical report, we developed a Generative Expert Metric System (\CMS) that accomplishes the user-specified task or goal specified in Section~\ref{sec:prob-def}. It can successfully match two teams $T_x$ and $T_y$ that the system identifies, where $T_x$ is identified as the team that most needs improvement with respect to the goal, and $T_y$ is the best fit for supporting $T_x$ toward achieving the goal. We detail how \CMS works, starting with an overview and followed by implementation details.

\subsection{System Overview}
\begin{figure}[htbp]
    \centering
    \includegraphics[width=\textwidth]{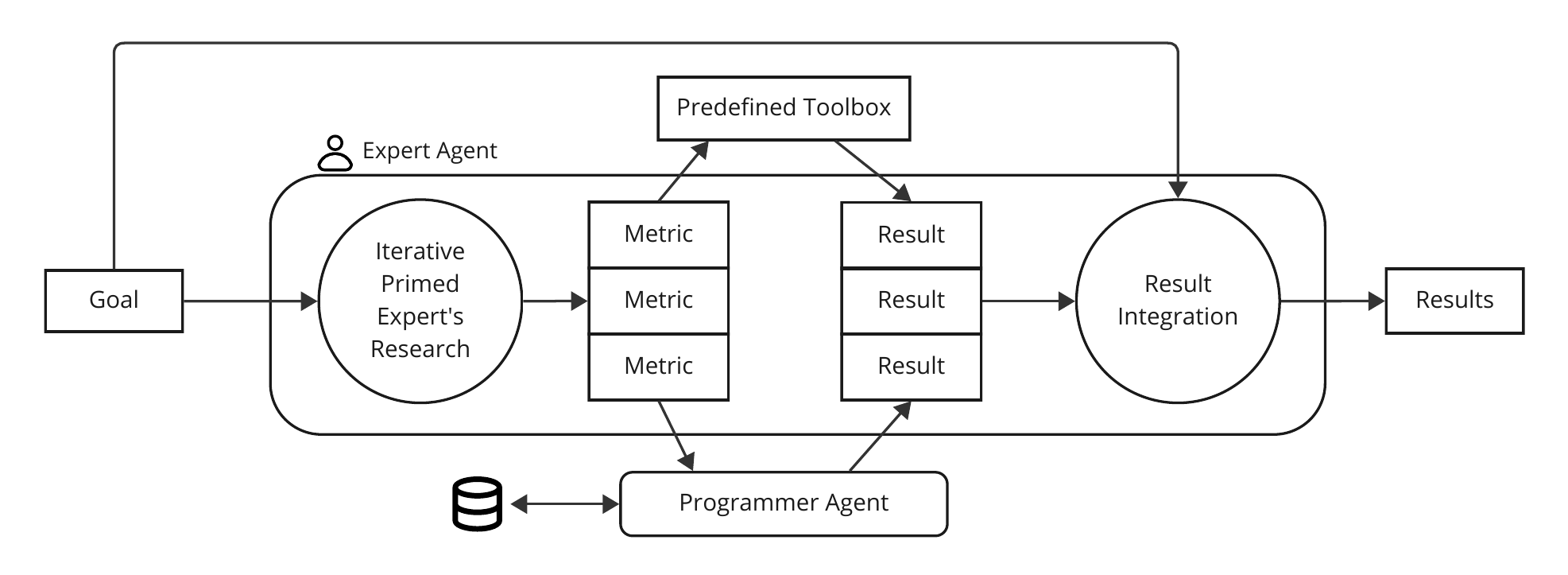}
    \caption{Expert Agents. The agents are `primed' by the given expert's research contributions to the field. This model is inspired by the generated knowledge prompting technique proposed by~\citet{liuGeneratedKnowledgePrompting2022}.}
    \label{fig:experts}
\end{figure}

To construct this system, we implemented multiple LFM-powered agents, orchestrated using an existing orchestration framework called AutoGen~\cite{wuAutoGenEnablingNextGen2023}. Additionally, we used Guidance~\cite{GuidanceaiGuidance2024}, the OpenAI API, and a MySQL database in the implementation. The database stores the data used to test our prototype, representing common information available to developer teams. We used software artifacts from open-source software (OSS) repositories including source code, commit details, and discussion threads. Each agent is designed and prompt-engineered to complete specific operations; see system overview in 
Figure~\ref{fig:overview} below. \CMS consists of the following components:
\vspace{1em}

\begin{description}[style=unboxed,leftmargin=0cm]
\setlength{\itemsep}{0pt}
\setlength{\parskip}{2pt}

\item[Lead Orchestrator] Acts as the primary interface and memory store, coordinating between the user, other agents, and the database to produce a final recommendation.
\item[Programmer] Crafts database queries to retrieve relevant information. Given a task, it designs proxies or refines a search based on its knowledge of the database.
\item[Expert] Leverages domain-specific knowledge of a human expert's research outcomes to make logical suggestions for \CMS. A breakdown of the expert agent is available in Figure~\ref{fig:experts}
\item[Judge] Employs well-established decision-making algorithms to finalize the team or community match, ensuring the recommendation aligns with the user intent specified.
\end{description}

\begin{figure}[htbp]
    \centering
    \includegraphics[width=\textwidth]{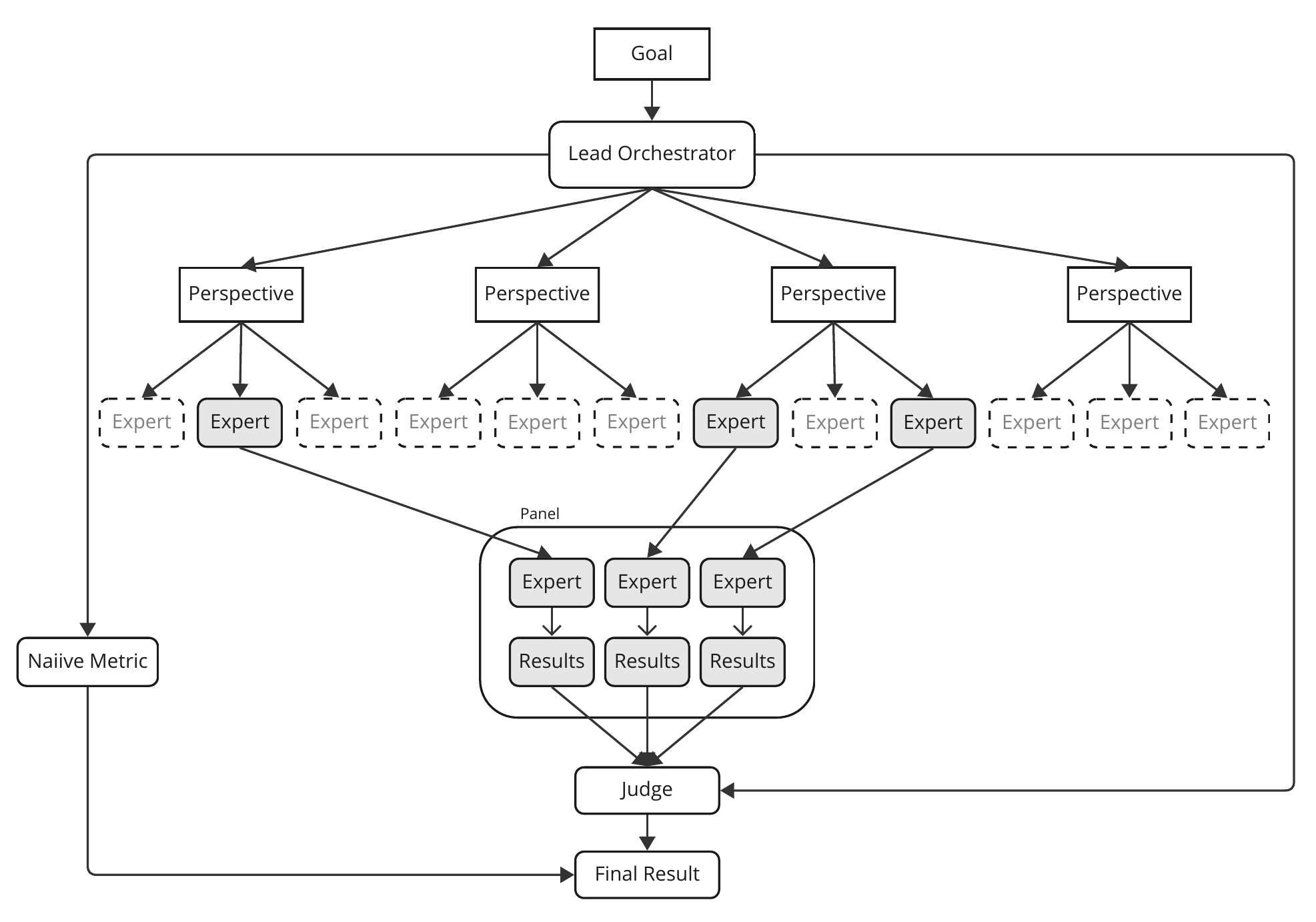}
    \caption{System Overview. The system begins with the Lead Orchestrator defining a naive metric to identify Team $T_x$. For simplicity, we omit the details of how this metric was generated. The Lead Orchestrator subsequently forms a panel of experts augmented through perspectives relevant to the given goal. The experts generate teams' results that pass through to a judge agent to form a decision. A final result is then aggregated.}
    \label{fig:overview}
\end{figure}

The scenario we used to drive the \CMS implementation is one where the user specifies the high-level goal G, e.g. improve developer on-boarding process, and expects team $T_x$ to be identified, followed by team $T_y$ that is the best-suited team or community match to drive $T_x$'s improvement with regard to goal G. Once the user submits the high-level goal to the \CMS,  the Lead Orchestrator proceeds with the following two tasks: 
\begin{enumerate}
    \item First, the Lead Orchestrator attempts to define a main goal metric (naively, i.e., without external information) that will serve as a proxy for the specified high-level goal and will be used to identify the target team ($T_x$) that is the worst performer with regard to the specified goal. The Lead Orchestrator presents this metric as a function name and a description of what it should measure.
    
    This information is then passed to a Programmer agent that will either select from a predefined set of functions available in \CMS that matches closest to the Lead Orchestrator's goal or construct a new function to implement this metric. The Programmer agent will return the resulting function to the Lead Orchestrator and the subsequently computed metric values for all teams within the database. Intuitively, this process is similar to a manager in a corporation asking each of their direct reports to provide a specific metric score for each team under their management to make an initial evaluation.

    \item Second, the Lead Orchestrator identifies relevant disciplines that provide insights toward the user-specified goal. Based on these disciplines, it generates and then judiciously assembles a panel of real-life experts in these disciplines, considering the panel expert selection criteria specified by the user, if any. The list of identified experts is shared with the user. Intuitively, this is a decision-maker assembling a panel of experts in relevant disciplines to tackle a complex problem.

\end{enumerate}

The \CMS system then asks each expert agent (a proxy for a real-life expert) to independently suggest several factors that could positively influence the main goal metric based on each identified domain expert’s knowledge. The factors identified by the expert agent as relevant to the matching decision are presented to the user.  

These factors are then passed to a Programmer agent as a function name and function description for each factor. The Programmer agent will either generate an implementation for this function on the fly, select from a list of predefined functions, or execute it for all teams in the database (minus team $T_x$). The expert agent will then collect the computed values and use them to generate a list of team rankings indicating the best team match(es) they think can positively influence the target team toward improving the target metric/goal. A short explanation generated by each expert agent, followed by the team rankings, will be passed back to the Lead Orchestrator. Intuitively, this phase identifies traits that can affect the main metric from different lenses and generates a list of potential team matches.

Finally, the \CMS system passes all the above information to the Judge agent. This agent takes into account any user-specified preference aggregation mechanism when aggregating the ranking information coming from the expert agents. Examples of such mechanisms include approval voting, 1 person 1 vote, or rank choice voting. The judging agent makes a decision and final selection, and the final team match is passed back to the Lead Orchestrator. This completes selecting the counterpart team $ T_y$ denoted in the pipeline. The final recommended team matching result is then presented to the user.

\subsection{Implementation Details}
In this subsection, we break down intuitions and provide implementation details for used across multiple expert agents and judges.

\subsubsection{Expert Agent and Iterative Prompt Priming}
Expert agents in~\CMS have two tasks to perform: 
\begin{enumerate}
    \item Design and define several metrics to measure and reflect on the given goal;
    \item Comprehend the results of these metrics to generate an interpretation of the aggregation of these results.
\end{enumerate}

To achieve Task 1, we developed and implemented a prompt engineering technique that we called \textbf{\textit{iterative prompt priming}} for expert agents. This process is inspired by~\cite {liuGeneratedKnowledgePrompting2022}, where knowledge is provided before answering a question to elicit more accurate responses. In the original method, the model is prompted to generate related pieces of knowledge before using this generated knowledge to answer the given question. This approach could not complete our goal, given the dimension and the complexity of our problem. Thus, we scaffold this methodology iteratively for expert agents to `prime' the language model to respond with more accurate and usable metrics. Our approach guides the knowledge-generation process with meta-prompting techniques that do not necessarily relate to the original question, only within context. We also integrated databases where expert agents can consume direct information. This iterative prompt priming process is guided through conversations, denoted as the first circle in Fig~\ref{fig:experts}. Since it is a conversation, the entire text from the previous stage is appended to the latter stage and passed to the language model. We detail them as three stages below:

\paragraph{\textbf{Stage 1}} The goal of stage 1 is to extract specific knowledge about the expert(s), absent information about the goal itself. The prompt requests detailed information regarding the experts' publications and insights into the field, together with a biography of the expert. This prompt-generated content refers to the expert's bibliography and describes the contributions made by the expert. One can easily imagine using a knowledge base or search tool to introduce specific content in this step that the LFM may not have access to, for example when including internal experts and their unique and specific expertise.  %

\begin{Verbatim}
    You are now consulting with {{expert}} in {{expert_field}}. Who is {{expert}}?
    Provide me with a bio of this expert.
    Then, discuss with {{expert}} on the following questions and more about {{expert}}:

      - What important publications and articles has {{expert}} written?
      - What are the insights from these publications and articles?

    This description should be extremely elaborate and detailed. 
    It should cover all the important aspects of {{expert}}'s work. 
    For each contribution, it should be a standalone paragraph that describes the contribution
    and insights.
  \end{Verbatim}

\paragraph{\textbf{Stage 2}} The goal of stage 2 is to prune unnecessary information generated in stage 1 by exposing the goal rather than a direct transformation for the goal. This selection process aims to prevent unrealistic or impractical proposals.

\begin{Verbatim}
    During this consultation with {{expert}}, you are now tasked to {{goal}} for team {{teamX}}. 
    Based on {{expert}}'s expertise, What do you think are important elements to consider
    when looking to {{goal}} for a team {{teamX}}? List out at least {{num_of_tools}} elements.
\end{Verbatim}

\paragraph{\textbf{Stage 3}} The goal of stage 3 is to transform the selected knowledge generated in stage 2 into metrics designed for this goal. This happens by exposing the kind of data available to the agent. The prompt gives further goal details and asks for proxies based on the previous elicitation with additional guidelines. The introduction of data scientists abstracts away the complexity of the system while creating function calls that are reasonable and quantitative with constraints. However, during this process, we do not ask for specifics regarding the implementation.

\begin{Verbatim}
    These elements make a lot of sense. Now, we need to consider the data we have. 
    {{db_description}}. You can use the table names to infer reasonable columns
    that are stored in the database with common knowledge. The list of table names 
    are provided here: {{db_tables}}. Based on the consultation above, discuss with {{expert}} 
    to find reasonable metrics that can be used to measure the elements you have listed out. 
    For each of the elements, you should try to come up with mutually exclusive metrics 
    that can be used to measure the elements. Remember, the expert's insights would not be helpful 
    if we cannot calculate a good metric to serve as a proxy using our dataset in the database.
    [cropped for brevity]
    ```
    <metric>
        <metric_name></metric_name>
        <metric_description></metric_description>
        <metric_reason></metric_reason>
    </metric>
    ```
\end{Verbatim}

We call this process iterative prompt priming because of the intentionally gradual exposure of the agent to the goal itself, guiding the model to process tokens based on the response from the previous stage. This unique approach directs the language model to generate content based on the expert's information. The subsequent steps guide the text exchange between the \CMS and the 'response' to better align with the content. During our design process, we found that guiding the LFM through seemingly out-of-context content toward the actual goal elicits better responses from individual expert agents than merely providing an expert's name or a direct inquiry. Although we lack direct statistical analysis to support this design choice, from a Natural Language Processing (NLP) perspective, we are intentionally weighting specific conditions for token generation in the latter process.
This design aligns with neuroscientific concepts, specifically predictive coding and synaptic plasticity. Predictive coding, as first published by ~\cite{raoPredictiveCodingVisual1999}, highlighted that the brain anticipates sensory inputs based on prior experiences, enabling vision neurons to process information more efficiently. Rather than asking an LFM-powered agent to provide information after being faced with a specific information elicitation task, the system leveraged this priming technique to establish the expert agent with prior knowledge in light of the complex problem. Synaptic plasticity refers to neural connections strengthening or weakening over time-based on activity levels, leading to more efficient learning and memory processes~\cite{kandelMolecularBiologyMemory2001}. The \CMS mirrors this process by selecting and engaging with a handful of expert agents, where each agent, primed with specific knowledge, represents a different activated neuron simulating the synaptic plasticity framework.

\subsubsection{Multiple Expert Agents}
~\CMS engages multiple independent expert agents with various perspectives to tackle a given goal. For example, team productivity can involve fields like organizational psychology, organizational behavior, and communication. As shown in Fig~\ref{fig:overview}, experts come in from different perspectives to produce a comprehensive response. A panel of experts is chosen to execute the algorithm described above.

This design aims to support the generalization of responses, which was motivated by the tree-of-thoughts~\cite{yaoTreeThoughtsDeliberate2023} approach. Although this implementation differs by using multiple experts following the same algorithm, rather than allowing for different reasoning processes, it allows diverse opinions, supporting a depth-first or breadth-first approach.

\subsubsection{Judges}
Judges incorporate human knowledge in designing decision-making mechanisms. Mechanism design influences the decision-making process and incentivizes different final results. For example, a plurality electoral rule tends to maintain dual-partisanship~\cite{rikerTwopartySystemDuverger1982}, while quadratic voting mechanisms aim to reduce the tyranny of the majority~\cite{posner2018radical}. 

\CMS allows users to specify at a high level how the ranking results should be aggregated without the need to know the specific aggregation mechanism. For instance, the user might want to request an `easy to understand and fair aggregation' which might cause ~\CMS to use the approval voting mechanism~\cite{bramsApprovalVoting2007}. The judge uses a ReAct framework~\cite{yaoReActSynergizingReasoning2023} which explains how it selected a specific decision-making mechanism to reduce errors. 

Once this mechanism is selected, the judge uses the algorithm of this mechanism in conjunction with the ranking results from all expert agents to derive the final decision.

\section{Methods}
We created $10$ different goals and generated metrics using \CMS and the OpenAI GPT-4 API without any prompt engineering. We conducted a qualitative comparative study to evaluate the metrics generated by \CMS. 

\paragraph{Goals}
We used ChatGPT-4 to generate five common challenges software development teams face. These topics were reviewed by co-authors on this paper. Given that this \CMS implementation was designed to support software engineering teams, we rewrote these challenges into goals that a software engineering manager might want to achieve. Understanding that not all goals could be easily expressed~\cite{holcomb2001asking}, we created abstract goals that transformed detailed descriptions into less specific requirements a software engineering manager might set out to achieve. We referred to those as~\textit{complex goals}  vs. ~\textit{abstract goals}. Here are two sets of examples:

\begin{itemize}
  \item Challenge: \textbf{Ensure Performance and Scalability}
    \begin{itemize}
      \item Complex Goal: Design and implement a scalable architecture that supports dynamic scaling and optimizes performance under varying load conditions.
      \item Abstract Goal: Build systems that grow with our success, handling more users effortlessly.
    \end{itemize}
  \item Challenge: \textbf{Enhance Cross-Team Coordination}
    \begin{itemize}
      \item Complex Goal: Establish a cross-functional liaison role responsible for coordinating activities and dependencies between teams, ensuring seamless integration and timely project progression.
      \item Abstract Goal: Create seamless collaboration across all teams to work as one unified force.
    \end{itemize}
\end{itemize}

\paragraph{Evaluation Process}

As \CMS allows the user to specify a series of parameters, we first describe the parameters used for this evaluation. We asked \CMS to generate metrics to achieve a certain goal by using the following parameters: 
\begin{enumerate}
    \item the number of perspectives or fields of study to consider when coming up with experts (i.e., 4)
    \item the number of experts to be selected for the panel of experts (i.e., 3)
    \item the expert selection criteria for the expert panel (i.e.,``"panel must contain 2 experts from the same field and 1 from a different field'')
    \item the number of metrics each expert will generate for use in the team matching process (i.e., 3). 
\end{enumerate}

More concretely, each time \CMS executed a task prompt based on the settings described above, the system generated 4 fields of study relevant to the specified task, then generated 3 experts for each field, as shown in Figure~\ref{fig:overview}. For example for a field of study like development operations, 3 real-world experts are generated: one known for a book about IT transformation, another for pioneering continuous delivery, and the third for impactful research on DevOps practices. A total of $12$ experts are generated, out of which 3 are selected for the expert panel that will then generate $9$ metrics total to be used for team ranking and final team matching. These ranking results are then aggregated to provide a final team match outcome. 

For comparison, we used OpenAI GPT-4 API to pass the same goals without specific prompt engineering. We call this the~\Vanilla approach, following works like~\citet{liuGeneratedKnowledgePrompting2022}. LangChain~\cite{LangChain} was used as a templating engine to generate the $9$ different metrics for the \Vanilla approach. We show the given prompt below:

\DefineVerbatimEnvironment{Verbatim}{Verbatim}{xleftmargin=5mm, frame=single, breaklines}
\begin{Verbatim}
  I am a software manager and I want my team to [{goal}]. I want to locate a different team in the company that can help me achieve this goal. What are metrics that I should look for, assuming I only have access to their GitHub repository. and assuming I have a data scientist that can help me craft specific queries for these metrics:
  ```
  [{{'function_name': 'function description'}}, ... ]
  ```
\end{Verbatim}

Both systems used the OpenAI GPT-4 API (version gpt-4-0613). It is important to note that not all cases passed to the \Vanilla version returned $9$ metrics and descriptions. We kept the descriptions as is to preserve the authenticity of our results. We did not cherry-pick results; metrics were generated only once, unless there was a rate limit error. After the metrics were generated, the first author reviewed the metric name and description generated by both systems, extracted them into a spreadsheet for comparison, and annotated the differences. These differences were then coded into cases shown in the next section.

\section{Results}\label{sec:findings}
In this section, we highlight the differences observed in the metrics generated by both systems---\CMS vs. \Vanilla. To illustrate key differences between systems, we summarize our findings and observations for each goal. Each of the 9 goals is denoted by a number, followed by \texttt{ABS} for abstract goal, or \texttt{CPX} for complex goal. We represent metrics generated by \CMS with \CMSIcon, and \Vanilla with \vanillaIcon. We reformatted the defined function names to \texttt{camelCase} for readability, and shared the descriptions generated as is for authenticity. Finally, we highlight observations from comparing the two systems with theoretical justifications for these differences.

\subsection{\CMS Generates Complex Metrics Built Through Compositions}

Metrics generated by \CMS often display greater complexity. Complexity here does not mean that the metric is more difficult to generate; instead, it indicates a composite function of direct metrics. In other words, it tries to compress different perspectives of measures into one measure. After matching the closest metrics from both systems, we show a few examples to highlight the difference. Table~\ref{tab:metric-complexity} presents three specific metric pairs across three different goals.

\begin{table}[h]
  \caption{Metric complexity across systems} \label{tab:metric-complexity}
  \arrayrulecolor{grayline}
  \begin{tabularx}{\textwidth}{>{\hsize=0.35\hsize}X>{\hsize=0.7\hsize}X}
  
  \multicolumn{2}{>{\hsize=\textwidth}X}{\textbf{Goal 02 CPX}: Implement a structured, cross-platform communication system enabling real-time collaboration and knowledge sharing among all software development teams.} \\ \hline
  \CMSIcon \texttt{userEngagement} & The number of comments, reactions, and reviews on issues and pull requests from each user over a defined time frame. \\
  \vanillaIcon \texttt{numberOfContributors} & The total number of contributors to the repository. \\ \hline
  \multicolumn{2}{>{\hsize=\textwidth}X}{\vspace{1px}\textbf{Goal 01 ABS} Adapt to market change, stay ahead of the curve, always ready to embrace change for our advantage.} \\ \hline
  \CMSIcon \texttt{rateOfChangeAdoption} & Measuring the frequency and volume of code changes or additions that have been accepted and merged into the base branch. \\
  \vanillaIcon \texttt{commitFrequency} & The frequency of commits. \\ \hline
  \multicolumn{2}{>{\hsize=\textwidth}X}{\vspace{1px}\textbf{Goal 05 ABS} Keep projects focused and on track, delivering what we promised on time.} \\ \hline
  \CMSIcon \texttt{trackAdaptivePlanning} & Identifies instances of reassignment of issues, changes in issue labels, or changes in milestone dates and computes the average frequency of these changes. \\
  \vanillaIcon \texttt{issueResolutionTime} & The time taken to resolve issues. \\ \hline
  \end{tabularx}
  \arrayrulecolor{black}
\end{table}

\vspace{.5em}
\noindent\textbf{Observations}: In the examples shown in Table~\ref{tab:metric-complexity}, \CMS generated more comprehensive and detailed metrics compared to \Vanilla. The \CMS metrics are more complex, as these metrics require mathematical operations or aggregation through composition. For example, \texttt{userEngagement} is a composition of various forms of user interaction over time using different actions. In contrast, the \Vanilla metric counts only the number of contributors without considering their level of activity or engagement. The \CMS metrics \texttt{rateOfChangeAdoption} and \texttt{trackAdaptivePlanning} demonstrate the multiple dimensions that \CMS elicits compared to the \Vanilla approach.

\subsection{\CMS Generates Metrics With Specificity Grounded In Theories}\label{sec:theories}
Metrics generated by \CMS often display specificity grounded in theoretical frameworks. Here, specificity does not indicate that the metric is more limited in scope; instead, it highlights a clear and focused measure grounded in established theories due to iterative prompt priming. Let us examine two examples of specificity in Table~\ref{tab:metric-specificity}.

\begin{table}[H]
  \caption{Metric specificity across systems} \label{tab:metric-specificity}
  \arrayrulecolor{grayline}
  \begin{tabularx}{\textwidth}{>{\hsize=0.35\hsize}X>{\hsize=0.7\hsize}X}
  
  \multicolumn{2}{>{\hsize=\textwidth}X}{\textbf{Goal 01 CPX} Implement an agile response strategy that allows the team to quickly pivot and adapt to emerging technologies and market demands.} \\ \hline
  \CMSIcon \texttt{refactoringFrequency} & This metric measures how frequently code refactoring is performed by examining the commit history for keywords related to refactoring or significant code alterations.[~\textellipsis{}] \\
  \vanillaIcon \texttt{codeChurn} & This function will measure the amount of code that is changed, added, or deleted over time. Agile teams should have low code churn rate as they aim for iterative and incremental development.\\ \hline
  \multicolumn{2}{>{\hsize=\textwidth}X}{\vspace{1px}\textbf{Goal 08 ABS} Build a team culture that celebrates success and fosters strong bonds among team members.} \\ \hline
  \CMSIcon \texttt{recognitionRatio} & [\textellipsis{}]the frequency of appreciative or congratulatory comments made by a user. This could be coupled with other positive emojis or phrases used to recognize others' work. [\textellipsis{}] \\
  \vanillaIcon \texttt{frequencyOfCommentsOnCommits} & Frequent comments can suggest a high level of collaboration and communication among team members. \\ \hline
  \end{tabularx}
  \arrayrulecolor{black}
\end{table}

\noindent\textbf{Observations}: In both examples, metrics are grounded in existing theories. In the first example, \CMS elected a renowned software developer, author, and speaker, recognized for contributions to software development practices and methodologies, as an expert to consult with. The corresponding metric that \CMS generated, \texttt{refactoringFrequency}, directly aligns with the importance of constant refactoring to support the software engineering team in adopting changes, as shown in prior research~\cite{fowlerRefactoringImprovingDesign2019}. Unlike the~\Vanilla metric~\texttt{codeChurn}, which tracks overall code updates, this metric specifically measures refactoring activities.

In the second example, \CMS was primed on an expert who brought emotional intelligence into the mainstream and integrated these concepts into leadership practices. Thus, unlike the~\Vanilla metric,~\CMS looks for specific cues such as '+1', 'heart', 'hooray', 'laugh', or 'rocket' in the comment as a proxy which aligns with emotional intelligence theories~\cite{goleman1998working}.

\begin{table}[H]
  \caption{Metrics operationalization across systems} \label{tab:metric-operation}
  \arrayrulecolor{grayline}
  \begin{tabularx}{\textwidth}{>{\hsize=0.35\hsize}X>{\hsize=0.7\hsize}X}
  
  \multicolumn{2}{>{\hsize=\textwidth}X}{\textbf{Goal 07 CPX} Form a technology review board that evaluates and recommends technologies based on current and future project needs, considering factors like scalability, maintainability, and team expertise.} \\ \hline
  \CMSIcon \texttt{calculateEconomicImpact} & \textellipsis{}calculate the 'economic impact' as the implementation cost minus the estimated revenue (ROI) that the technology would generate. The implementation cost can be inferred from (the) 'investment data' table, and the ROI could be estimated based on past performance data for similar technologies under 'repository history' and 'technology usage' records. \\
  \vanillaIcon \texttt{N/A} & Nothing comparable or of similarity\\ \hline
  \multicolumn{2}{>{\hsize=\textwidth}X}{\vspace{0.5px}\textbf{Goal 09 ABS} Become the best place to work for software developers by offering growth and stability.} \\ \hline
  \CMSIcon \texttt{measureAutonomy} & \textellipsis{}the autonomy level of a software developer~\textellipsis{} include factors such as how often they're allowed to choose their own tools, set their own deadlines, or decide their approach to problem-solving.~\textellipsis{}\\
  \vanillaIcon \texttt{N/A} & Nothing comparable or of similarity\\ \hline

  \end{tabularx}
  \arrayrulecolor{black}
\end{table}

\subsection{\CMS Generates Metrics Harder to Operationalize and With More Assumptions}
\CMS metrics offer valuable, holistic insights, but also bring challenges. First, the metric complexity and specificity can make them harder to operationalize and implement automatically. Second, some metrics come with specific assumptions that are not apparent from their descriptions. We illustrate these differences by comparing \Vanilla and \CMS metrics in Table~\ref{tab:metric-operation}.

\vspace{.5em}
\noindent\textbf{Observations}: In Table~\ref{tab:metric-operation}, \CMS generated two metrics that reflect concepts stemming from existing literature which \Vanilla did not generate. However, both metrics cannot be derived solely based on code repository information or readily available data. While the descriptions pointed to possible proxies that might be able to estimate these values,~\CMS made assumptions that information such as the implementation cost of a technology is available to the system, or to the individuals operating the tool. This makes these metrics difficult to scale and make use of. That being said, these metrics are the first steps for teams to locate reasonable measures as proxies that reflect them.

\subsection{\CMS Generates More Diverse Metrics}
In spite of the occasional metric that is hard to interpret and operationalize, the metrics generated by \CMS exhibit greater diversity. The \Vanilla system tends to repeat or use similar metrics, despite the wide range of goals specified. For example, across the 178 metrics generated by the \Vanilla system for $10$ cases, the top $5$ frequently identified metrics are: \texttt{issueResolutionTime} appearing $14$ times, \texttt{commitFrequency} appearing $13$ times, \texttt{testCoverage} appearing $12$ times, \texttt{codeChurn} appearing $11$ times, \texttt{contributorsCount} appearing $11$ times. These $5$ metrics account for 34\% of all 178 metrics.

On the other hand, \CMS generated 180 metrics with the following top $5$ frequently identified metrics: \texttt{calculateAverage}-\texttt{IssueResolutionTime} appearing $5$ times, \texttt{communicationFrequency} appearing $4$ times, \texttt{measureTestCoverage} appearing $3$ times, \texttt{calculateCollaborationScore} appearing $3$ times, and \texttt{alignmentScore} appearing $3$ times. These metrics account for only 10.11\% percent of all $180$ metrics.

\begin{landscape}
  \begin{table}[h]
    \centering
    \caption{Goal 05 ABS: Keep projects focused and on-track, delivering what we promised on time. \codeIcon denotes metrics that were successfully automatically implemented.}
    \label{tab:g05abs}
    \arrayrulecolor{grayline}
    \begin{minipage}[t]{0.67\textwidth}
      \small
      \subcaption{Metrics generated by \CMS}
      \label{tab:cms-metrics}
      \arrayrulecolor{grayline}
      \begin{tabularx}{\textwidth}{>{\hsize=0.2\hsize}X>{\hsize=0.8\hsize}X}
        \hfill\CMSIcon & Metrics generated by \CMS \\ \hline
        \codeIcon\texttt{Calculate}-\texttt{Incremental}-\texttt{DeliveryRate} & [\textellipsis{}] the number of closed issues or merged pull requests over a specific time period [\textellipsis{}] timestamp and statuses of issues and pull requests to measure this. \\ \hline
        \codeIcon\texttt{Compute}-\texttt{Communication}-\texttt{Efficiency} & [\textellipsis{}] average time it takes from the moment a new issue is created, until the first comment is made or action is taken. It will infer this information from the timestamp of issue creation and the timestamp of first interaction (comment or action). \\ \hline
        \codeIcon\texttt{Track}-\texttt{Adaptive}-\texttt{Planning} & [\textellipsis{}] how quickly and frequently the project's plans or priorities are adapted in response to new developments or changes. It identifies instances of reassignment of issues, changes in issue labels, or changes in milestone dates, and computes the average frequency of these changes. \\ \hline
        \codeIcon\texttt{Feature}-\texttt{Complexity}-\texttt{Index} & [\textellipsis{}] average number of features (or tasks) per release (or milestone). This calculation will require access to the 'milestone' and 'issue' tables to count the number of issues completed per milestone. Additionally, dependencies and relations between issues can also be used to infer complexity. \\ \hline
        \codeIcon\texttt{Mean}-\texttt{Time}-\texttt{To}-\texttt{Resolve} & [\textellipsis{}] the average time taken to close an issue once it is opened. It will need data from the 'issue' table to calculate the time difference between the creation and closing of each issue. \\ \hline
        \codeIcon\texttt{Task}-\texttt{Completion}-\texttt{Rate} & [\textellipsis{}] ratio of tasks or issues that are closed to the total number of tasks. This calculation requires access to the 'issues' table to obtain data about created and closed issues. \\ \hline
        \codeIcon\texttt{Calculate}-\texttt{Average}-\texttt{Issue}-\texttt{Resolution}-\texttt{Time} & [\textellipsis{}] average time taken to resolve issues. It would involve fetching data from tables storing issue creation and closing timestamps and calculating the difference in days. \\ \hline
        \codeIcon\texttt{Calculate}-\texttt{Backlog}-\texttt{Management}-\texttt{Efficiency} & [\textellipsis{}] the time taken (in days) between adding an issue to the backlog and that issue being closed. It would involve extracting data from the tables that store issue details and timestamps of when issues were created and closed. \\ \hline
        \codeIcon\texttt{Calculate}-\texttt{Sprint}-\texttt{Completion}-\texttt{Rate} & [\textellipsis{}] the percentage of tasks (issues and pull requests) from the backlog that were completed in a sprint, inferred by open and closed timestamps and commit information. It would require data from the tables that store issue, pull request and commit details, appropriately joined and filtered by time frames. \\ \hline
      \end{tabularx}
      \arrayrulecolor{black}
    \end{minipage}%
    \hspace{0.3em}
    \begin{minipage}[t]{0.58\textwidth}
      \small
      \subcaption{Metrics generated by \Vanilla}
      \label{tab:vanilla-metrics}
      \arrayrulecolor{grayline}
      \begin{tabularx}{\textwidth}{>{\hsize=0.2\hsize}X>{\hsize=0.8\hsize}X}
        \hfill\vanillaIcon & Metrics generated by \Vanilla \\ \hline
        \texttt{Bug}-\texttt{Fixing}-\texttt{Rate} & [\textellipsis{}] the rate at which bugs are fixed within the repository. A high bug fixing rate signifies a team that is effective at identifying and addressing issues. \\ \hline
        \texttt{Code}-\texttt{Churn} & [\textellipsis{}] amount of code that gets modified or deleted after being checked in, which is an indicator of code stability and can suggest the effectiveness of the team. \\ \hline
        \texttt{Code}-\texttt{Quality}-\texttt{Metrics} & [\textellipsis{}] quality of the code in the repository. High-quality code indicates a team that values accuracy and efficiency. \\ \hline
        \texttt{Code}-\texttt{Review}-\texttt{Time} & [\textellipsis{}] time taken to review and merge code. A shorter review time suggests a team that can efficiently manage its workload. \\ \hline
        \texttt{Commit}-\texttt{Frequency} & [\textellipsis{}] frequency and regularity of commits. A team that regularly contributes to their repository is likely to be more focused and able to deliver on time. \\ \hline
        \texttt{Contribution}-\texttt{Diversity} & [\textellipsis{}] diversity of contributions within the repository. A diverse range of contributions suggests a team with a wide range of skills and expertise. \\ \hline
        \texttt{Issue}-\texttt{Resolution}-\texttt{Time} & [\textellipsis{}] the time taken to resolve issues. A shorter resolution time indicates a more efficient team. \\ \hline
        \texttt{Project}-\texttt{Completion}-\texttt{Rate} & [\textellipsis{}] rate at which projects are completed within the repository. A high project completion rate indicates a team that can deliver what it promises on time. \\ \hline
        \texttt{Pull}-\texttt{Request}-\texttt{Rate} & [\textellipsis{}] rate at which pull requests are created and closed. A high pull request rate may signify an active team. \\ \hline
      \end{tabularx}
      \arrayrulecolor{black}
    \end{minipage}
  \end{table}
\end{landscape}

\begin{landscape}
  \begin{table}[h]
    \centering
    \caption{Goal 05 CPX: Implement a rigorous project management protocol that requires detailed documentation and approval for all changes to project scope or objectives. \codeIcon denotes metrics that were successfully automatically implemented.}
    \label{tab:g05cpx}
    \arrayrulecolor{grayline}
    \begin{minipage}[t]{0.67\textwidth}
      \small
      \subcaption{Metrics generated by \CMS}
      \label{tab:cms-metrics-g5cpx}
      \arrayrulecolor{grayline}
    \begin{tabularx}{\textwidth}{>{\hsize=0.2\hsize}X>{\hsize=0.8\hsize}X}
      \hfill\CMSIcon & Metrics generated by \CMS \\ \hline
        \texttt{calculate}-\texttt{Risk}-\texttt{Identification}-\texttt{And}-\texttt{Plan} &  [\textellipsis{}] identification of risks and their mitigation planning related to the changes in the project scope. This can be done by analyzing the issue tracking data and corresponding mitigation steps taken, and whether the issues were associated with scope changes in the git commits or pull requests.\\ \hline
        \texttt{calculate}-\texttt{Scope}-\texttt{Change}-\texttt{Frequency} &  [\textellipsis{}] number of times the scope/objective of the project changes. [\textellipsis{}] counting the number of times significant modifications are made to the git repository, indicated by either major commits or pull requests that significantly alters the codebase or its documentation.\\ \hline
        \codeIcon\texttt{measure}-\texttt{Team}-\texttt{Involvement}-\texttt{In}-\texttt{Changes} & [\textellipsis{}] involvement of team members in making changes to the project scope or objectives. [\textellipsis{}] tallying the number of unique individuals contributing to commits or pull requests associated with significant changes, along with the frequency of their contributions.\\ \hline
        \texttt{impact}-\texttt{Analysis}-\texttt{Score} & [\textellipsis{}] calculated by assessing the proportion of proposed changes that were backed by an impact analysis report. An impact analysis report documents the anticipated effects of the proposed change on other aspects of the project.\\ \hline
        \texttt{process}-\texttt{Adherence}-\texttt{Score} & [\textellipsis{}] comparing the number of changes made in a project that followed the documented process against the total number of changes made. [\textellipsis{}] whether the proposal, documentation, and approval steps were correctly undertaken for every change made.\\ \hline
        \codeIcon\texttt{team}-\texttt{Communication}-\texttt{Score} & [\textellipsis{}] analyzing the number of documented communications (like discussions, meetings records, or emails) related to a proposed change against the total number of changes made. A higher ratio indicates a more robust communication framework in place. \\ \hline
        \codeIcon\texttt{change}-\texttt{Approval}-\texttt{Rate} &  [\textellipsis{}] proportion of proposed changes to the project scope or objectives that went through [\textellipsis{}] rigorous project management protocol. This can be measured by checking if a formal approval record exists for each proposed change. \\ \hline
        \texttt{documentation}-\texttt{Compliance}-\texttt{Score} &  [\textellipsis{}] calculates a score based on the completeness and timeliness of the documentation updates. Completeness can be gauged by checking if all required sections of the documentation are filled out. Timeliness can be checked by seeing if the documentation update was done within a pre-specified time [\textellipsis{}] (like change approval).\\ \hline
        \texttt{stakeholder}-\texttt{Involvement}-\texttt{Index} & [\textellipsis{}] level of stakeholder involvement in decision-making processes. [\textellipsis{}] tracking the frequency and degree of engagement of stakeholders in important discussions, their involvement in the review and approval of proposed changes, and their presence in meetings.\\ \hline
      \end{tabularx}
      \arrayrulecolor{black}
    \end{minipage}%
    \hspace{0.3em}
    \begin{minipage}[t]{0.58\textwidth}
      \small
      \subcaption{Metrics generated by \Vanilla GPT4API}
      \label{tab:vanilla-metrics-g5cpx}
      \arrayrulecolor{grayline}
    \begin{tabularx}{\textwidth}{>{\hsize=0.2\hsize}X>{\hsize=0.8\hsize}X}
        \hfill\vanillaIcon & Metrics generated by \Vanilla \\ \hline
        \texttt{code}-\texttt{Reviews} & The presence of code reviews shows a commitment to quality assurance, a key component of rigorous project management. \\ \hline
        \texttt{commit}-\texttt{Messages} & Analyzing commit messages can provide insights into how well the team documents their changes. Look for detailed, meaningful messages. \\ \hline
        \texttt{issues}-\texttt{Created}-\texttt{And}-\texttt{Closed} & This can show how well the team manages their project's scope and tracks bugs and desired features. \\ \hline
        \texttt{merge}-\texttt{Conflicts} & A low number of unresolved merge conflicts may indicate a team that effectively manages their workflow. \\ \hline
        \texttt{number}-\texttt{Of}-\texttt{Branches} & Frequent use of branches may indicate that the team is using a feature branching workflow, often associated with rigorous project management protocols. \\ \hline
        \texttt{number}-\texttt{Of}-\texttt{Commits} & This will show you how active the team is. A team that's well-versed in project management should have a consistent number of commits. \\ \hline
        \texttt{number}-\texttt{Of}-\texttt{Contributors} & A higher number of contributors may indicate a larger or more collaborative team. \\ \hline
        \texttt{pull}-\texttt{Requests} & The number of pull requests, their frequency, and the quality of their descriptions can also indicate how rigorously the team manages their work. \\ \hline
        \texttt{release}-\texttt{Frequency} & The frequency and regularity of releases could indicate a well-structured development process. \\ \hline
      \end{tabularx}
      \arrayrulecolor{black}
    \end{minipage}
  \end{table}
\end{landscape}

\subsection{Detailed Case Analysis}
\label{sec:case_sample}
In the previous subsection, we discussed characteristics observed across all the metrics generated for the $10$ cases by selecting pairs of metrics between the two systems. In this subsection, we focus on two specific goals and describe these differences holistically. We selected~\textit{Goal 05}, relating to the challenge of ``Prevent Scope Creep'' that many software developer teams face. The two derived goals are listed as:

\begin{itemize}
  \item Goal 05 ABS: Keep projects focused and on-track, delivering what we promised on time.
  \item Goal 05 CPX: Implement a rigorous project management protocol that requires detailed documentation and approval for all changes to project scope or objectives.
\end{itemize}

We select~\textit{Goal 05} for the detailed case analysis because the programmer agent in~\CMS was able to automatically generate the code implementation (using AutoGen) for all the metrics proposed by~\CMS for~\textit{Goal 05 ABS}. Thus, for this goal we show a fully functional, end-to-end process of metrics being generated, implemented, and executed to find a team match that achieves the given goal. The other goals that did not have code implementations for all metrics successfully auto-generated were not marked as failures since automatic code generation is not the primary goal of this paper. We limit the scope of this project to metric generation because code generation itself is a separate, growing field of study~\cite{beurer2023prompting, wuAutoGenEnablingNextGen2023, kim2020natural}. This example provides a baseline benchmark to compare against for future implementations, as more advanced self-coding systems can only enhance \CMS performance. Table~\ref{tab:g05abs} and Table~\ref{tab:g05cpx} list results for~\textit{Goal 05 ABS} and~\textit{Goal 05 CPX}, respectively. We placed a ~\codeIcon for the metrics where code implementation was auto-generated by AutoGen, and then executed by~\CMS by using software communities data from the database. For example, for the first metric in Table~\ref{tab:cms-metrics-g5cpx}, AutoGen generated Python and SQL code that implements ~\texttt{CalculateIncrementalDeliveryRate} by executing SQL code querying the database for the number of closed issues in $2022$.

When comparing metrics generated by~\CMS and the~\Vanilla approach, as shown in Table~\ref{tab:g05abs}, we see that they echo the characteristics discussed in the previous subsection. When we consider the $9$ metrics \CMS generated, we see that
some metrics can be similar in terms of the attributes they consider. For instance, \CMS generated~\texttt{MeanTimeToResolve} and~\texttt{ComputeCommunicationEfficiency}. The former calculates the average time it takes for an issue to be resolved. The latter considers the average time an issue receives an interaction. As such, both metrics implicitly prioritize and weight specific measures more, in this case, time and issues, by creating variations within all $9$ metrics that make use of these same measures.

Other times, composite metrics that \CMS generated contain measures that other metrics cover. In this example, \texttt{calculateSprintCompletionRate} partially covers the ratio of open to closed issues which \texttt{taskCompletionRate} aims to measure. This overlap became an implicit weighting of specific measures when the final judge considered all $9$ metrics. 

In contrast with \CMS metrics, metrics generated by the \Vanilla approach are broad and encompass a wide variety of items that do not necessarily address the given goal. The metrics typically represent general software engineering concepts without being tailored specifically to the given task. 

Next, we examine how metrics change when goals become more specific. Shifting from abstract to concrete, we compare the metrics listed in Table~\ref{tab:g05cpx}. First, we notice the sensitivity in metrics generated by \CMS in this case. Adding concepts such as ``project management protocol'' and ``detailed documentation'' to the goal specification led to the increase in metrics that are less technical, such as \texttt{ProcessAdherenceScore} and \texttt{calculateRiskIdentificationAndPlan}, as well as specific measures such as \texttt{documentationComplianceScore}. While~\CMS did not automatically generate the corresponding code for these metrics, implementations for them could be crafted with human intervention, or using more advanced data analysis tools. These metrics, on the other hand, highlight better metrics that can contribute to realizing the given goal. On the other hand, when we look at the vanilla metrics generated for the complex goal, the metrics generated remain low-level snapshots of repository information. Furthermore, many of these metrics are very similar to the metrics generated using the abstract goal.

In summary, when considering Goal $05$, we showed that~\CMS generated $9$ metrics that implicitly weighted specific measures that it believed were critical to improving the given goal. When comparing metrics between the abstract and complex goals, we observe that~\CMS generated metrics are more sensitive to the change and tailored the responses by using specific items grounded in theories.

\section{Discussion}\label{sec:discussion}

\subsection{\CMS as a Copilot}
\CMS acts as a copilot in constructing proxies tailored to the specified goal, and using the data at hand. As noted in the Results section, the system sometimes generates metrics that are overly narrow, or overfits its conclusions. As such \CMS plays the role of the real-world experts who bring~\textbf{theoretical insights} to the given problem, and local individuals familiar with the~\textbf{context of the problem} would guide the experts to apply these insights effectively.

The literature on copilot roles and expert-novice collaboration supports this. Constructing knowledge requires active participation, developed and used through engagement with the activity, context, and culture~\cite{laveSituatingLearningCommunities1991}. \CMS lowers the barriers by surfacing expert knowledge directly to individuals without actual experts present. In the example, the metric~\texttt{calculateEconomicImpact} shown in Table~\ref{tab:metric-operation} encouraged using the ROI of previous technology usages as a proxy for evaluating technologies out of thousands, if not millions, of possible metrics. As models improve with advancements in code generation and logical reasoning, they will be able to offer better, more specific recommendations. However, individuals familiar with the problem context will still need to guide metrics, positioning \CMS as a Copilot rather than a people replacement.

\subsection{Mapping to Expert Decisions}

One of the strengths of the metrics generated by \CMS is its composition and specificity; akin to the work of real-world experts. 
For instance, experts often already had a sense of the specific relationships, concepts, and thresholds among a selection of measurements. \CMS takes a similar approach where measurements then form metrics when a scenario is given.

Herbert Simon's work on intuition and judgment posits that intuition and sound judgments are outcomes of extensive practice and experience, which he denotes as ``frozen analysis''~\cite{simon1997models}. These internalized analyses allow experts to respond rapidly to familiar patterns when given a context. In many cases, experts rely on experience-based intuition and pattern recognition in dynamic environments~\cite{kleinDecisionMakingAction1993}, thereby reducing errors~\cite{hammond1987direct}. These findings support the use of the iterative prompt priming technique that \CMS employs. 

By making pre-processed information (in this case an expert's prior works) available to the Large Foundation Model (LFM), the model does not have to regenerate or digest information from scratch, allowing the model to generate more accurate and expert-like decisions. Hence, we observe the different metrics generated by the~\Vanilla model and the \CMS system.

Recall the metrics grounded in existing theories listed in Sec.~\ref{sec:theories}. If \CMS was able to map critical works in the field of experts who have unique, significant experiences to specific metrics, it is reasonable to project its capability when users within a software company provide \CMS with information available internally only. This information can be knowledge bases, personas, or even experts who do not have an external facing profile, but have significant know-how or skill sets within the organization. In these settings, \CMS would be able elect and translate these characteristics into concepts that inform context-dependent metrics. This tool has the potential to transform how individuals within organization solicit the unique expertise of their colleagues and evolve their approach to collaboration.

\subsection{Blueprinting Metrics to Solve Difficult Problems using LFMs}

\CMS, in its role as a copilot, demonstrates its ability to forge expert knowledge into context-specific solutions, effectively crafting blueprints for solving difficult problems by leveraging the capabilities of Large Foundation Models (LFMs).

Prior research highlights two primary challenges when it comes to crafting metrics. The first challenge lies in initiating metrics that can effectively measure and resolve issues~\cite{maoHowDataScientistsWork2019, grayMeasurementMadnessRecognizing2014}. Once this challenge is overcome, teams may still misuse inappropriate metrics for their goals. For example, in software engineering, developers frequently equate productivity with lines of code, a simplistic measure that fails to capture the true complexity of team output. Experts have corrected such measurements by developing metrics that better reflect a team's collective contributions, encompassing multiple dimensions of developer work~\cite{forsgrenSPACEDeveloperProductivity2021}. In these situations, teams typically require expert guidance to navigate and select metrics \textit{suitable} for comparison.

\CMS positions itself as a blueprinting tool, guiding the creation of a more comprehensive set of metrics. These blueprints are designed to stimulate informed discourse among teams, leading to better comparisons and decisions without an immediate need for physical experts, particularly when teams face complex and challenging problems.

The iterative prompt priming technique with the expert panel selection design translates complex problems into multiple perspectives. Thus, even if \CMS introduces complex and less operationalized metrics, it prevents novices from selecting more accessible but possibly poor metrics. Instead, it seeds a conversation that overcomes this hurdle, allowing local expertise and discussion to facilitate more situated metrics based on \CMS's initial recommendations. This design also allows CMS to generate metrics covering both breadth and depth. The case study in the results section (Sec.~\ref{sec:case_sample}) demonstrates the breadth of metrics generated, showcasing the system's ability to identify experts from diverse domains. While priming in behavioral economics can introduce biases, intentionally priming models are used here to reduce overly narrow metrics and solutions.

\subsection{Limitations and Future Work}
While~\CMS metrics show that such systems can generate~\textit{better} metrics given a goal and a dataset, we do not know how individual users would use such a system. Our report did not evaluate how decision-makers use team pairings generated by GEMS. The system still relies on users providing authentic and trustworthy information. Correctness and transparency in software version control are crucial for accurate pairing suggestions. Decision-makers must take responsibility for final decisions and effectively use systems like \CMS. Thus, future work should experiment with managers and decision-makers to understand how they evaluate the metrics generated and make use of systems like~\CMS.

\section{Conclusion}
Identifying suitable metrics to address challenging goals in improving software engineering practices is inherently difficult. Traditionally, individuals have often relied on suboptimal metrics that are more readily available. While prior research suggests the possibility of learning tacit knowledge from contextually similar teams, it still necessitates the design of measures that can effectively identify these teams.

In this technical report, we proposed~\CMS as a prototype applied to OSS community data powered by Large Foundation Models (LFMs) to assist individuals in designing more suitable metrics for specific software practice challenges. We introduced the iterative prompt priming technique and designed a multi-agent expert panel with a judge in our implementation. Our results highlight the differences between the metrics generated by \CMS and those produced by a~\Vanilla question-and-answer approach when addressing a given goal, illustrating the potential impact of~\CMS in a software engineering industry.

Through our case analyses, we demonstrated that \CMS generates complex metrics grounded in theories, making them more specific to the given problem. Additionally, \CMS produces a more diverse set of metrics that adhere more closely to the specified goals. Finally, we showed that \CMS-generated metrics are more sensitive to the nuances of specific goals.

Our findings have significant implications for how LFMs can be utilized to obtain expert perspectives, particularly when direct access to experts is limited. We envision future work that further investigates software engineering managers' and decision-makers' perceptions when provided with metrics generated by \CMS.

\begin{acks}
We would like to thank our Microsoft colleagues, interns, and partners who provided feedback that helped guide the direction of this work. Ti-Chung Cheng conducted this work as a research intern in Microsoft Research's Software Analysis and Intelligence in Engineering Systems (SAINTES) Group.
\end{acks}

\bibliography{tcheng}

\appendix
\section{Supplementary Diagrams}\label{app:supp-diagrams}

\begin{figure}[h]
    \centering
    \includegraphics[width=0.8\linewidth]{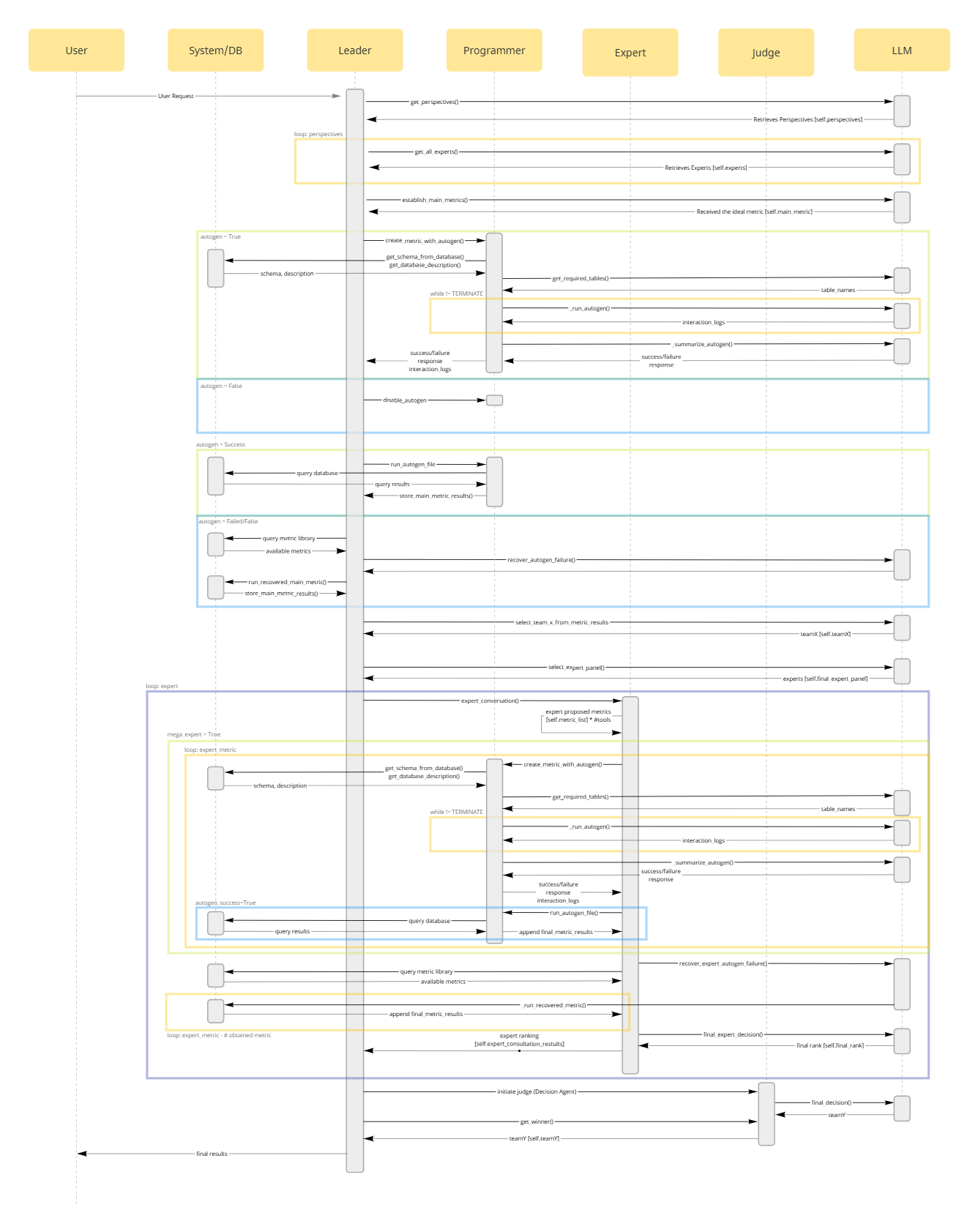}
    \caption{System Architecture Diagram of \CMS}
    \label{fig:sys-arch}
\end{figure}

\section{Expanded Reference Cases}
We provide the remaining $4$ cases in the appendix.

\begin{landscape}
  \begin{table}[h]
    \centering
    \caption{Goal 01 ABS: Stay ahead of the curve, always ready to embrace change for our advantage. \codeIcon denotes metrics that were successfully automatically implemented.}
    \label{tab:g01abs}
    \arrayrulecolor{grayline}
    \begin{minipage}[t]{0.67\textwidth}
      \small
      \subcaption{Metrics generated by \CMS}
      \label{tab:g01abs-cms-metrics}
      \arrayrulecolor{grayline}
      \begin{tabularx}{\textwidth}{>{\hsize=0.2\hsize}X>{\hsize=0.8\hsize}X}
        \hfill\CMSIcon & Metrics generated by \CMS \\ \hline
        \texttt{CountModern}-\texttt{MethodUsage} & [\textellipsis] the usage of modern software development methodologies in a repository. This can be measured by looking at the commit messages for references to methodologies such as Agile, DevOps, or other practices. A higher count suggests a more modern approach to software development. \\ \hline
        \codeIcon\texttt{Measure}-\texttt{Learning}-\texttt{Adaptation} & [\textellipsis] the adoption of new technologies or methodologies over time. This might be inferred through the introductions of new libraries or tools in the repository, significant shifts in coding practices, or introductions of new types of issue labels. A team that quickly adopts new technologies is likely to be proactive in staying current with new developments. \\ \hline
        \codeIcon\texttt{Track}-\texttt{Innovation}-\texttt{Creativity} & [\textellipsis] the number and frequency of innovative, non-traditional solutions adopted in a repository. This can be evaluated by assessing the implementation of unique coding practices or pioneering technologies in a repository, as well as the number of forks or stars a repository gets. High numbers suggest a high degree of innovation and creativity. \\ \hline
        \codeIcon\texttt{frequencyOf}-\texttt{IssueOpened} & [\textellipsis] rate at which new issues are opened for a given repository in a given period. This can be calculated using the count of new issues created divided by the total time observed. \\ \hline
        \texttt{numberOfActive}-\texttt{Contributors} & This metric measures the number of unique active contributors to a particular repository within a defined timeframe. This can be computed by tracking the unique users making commits. \\ \hline
        \codeIcon\texttt{rateOf}-\texttt{ChangeAdoption} & [\textellipsis] rate at which changes (like new features or bug fixes) are adopted in a particular repository. This can be calculated by measuring the frequency and volume of code changes or additions that have been accepted and merged to the base branch. \\ \hline
        \texttt{AgilePractice}-\texttt{AdoptionMetric} & [\textellipsis] frequency of commit histories and pull requests, the essential elements of Agile practices. It calculates the average number of commits and pull requests per week for each repository. Higher frequency suggests a more Agile and iterative code development practice. \\ \hline
        \codeIcon\texttt{Code}-\texttt{Refactoring}-\texttt{Metric} & [\textellipsis] frequency of code refactoring by counting the average number of code modifications per commit. Frequent modification indicates code optimization or refactoring exercises are taking place. \\ \hline
        \texttt{Innovation}-\texttt{Experimentation}-\texttt{Metric} & [\textellipsis] evaluates the extent of innovation and experimentation by monitoring the introduction of new tools and technologies in the repositories, inferred by the frequency of files with new extensions and the introduction of new topics in the repository. \\ \hline
      \end{tabularx}
      \arrayrulecolor{black}
    \end{minipage}%
    \hspace{0.3em}
    \begin{minipage}[t]{0.58\textwidth}
      \small
      \subcaption{Metrics generated by \Vanilla}
      \label{tab:g01abs-vanilla-metrics}
      \arrayrulecolor{grayline}
      \begin{tabularx}{\textwidth}{>{\hsize=0.2\hsize}X>{\hsize=0.8\hsize}X}
        \hfill\vanillaIcon & Metrics generated by \Vanilla \\ \hline
        \texttt{codeChurn} & [\textellipsis] the amount of code that is added and then removed or changed. Lower code churn can indicate stable, mature codebase while higher code churn may indicate active development and willingness to experiment with changes. \\ \hline
        \texttt{codeReviewTime} & [\textellipsis{}] the time taken to review and merge pull requests. A shorter review time can indicate efficiency and readiness to move forward with changes. \\ \hline
        \texttt{commit}-\texttt{Frequency} & [\textellipsis{}] the frequency of commits. Frequent commits can indicate the team's continuous effort to improve and adapt. \\ \hline
        \texttt{contributor}-\texttt{Count} & [\textellipsis{}] the number of contributors. A high number of contributors can indicate a diversified team that can bring in different perspectives for change. \\ \hline
        \texttt{dependency}-\texttt{Update}-\texttt{Frequency} & [\textellipsis{}] how often dependencies are updated. Regular updates indicate the team's readiness to adapt to new technologies and changes. \\ \hline
        \texttt{issuesOpened}-\texttt{ClosedRatio} & [\textellipsis{}] the ratio of opened to closed issues. A ratio close to 1 can indicate the team's ability to resolve issues and adapt to changes. \\ \hline
        \texttt{mergedPull}-\texttt{Requests} & [\textellipsis{}] the number of pull requests that have been merged. A high number of merges can indicate the team's ability to successfully implement changes. \\ \hline
        \texttt{pullRequest}-\texttt{Frequency} & [\textellipsis{}] the frequency of pull requests. High frequency can indicate that the team is continuously working on improving their codebase. \\ \hline
        \texttt{release}-\texttt{Frequency} & [\textellipsis{}] how often new releases are made. Frequent releases can indicate a team that's able to quickly respond to changes and deliver results.  \\ \hline
      \end{tabularx}
      \arrayrulecolor{black}
    \end{minipage}
  \end{table}
\end{landscape}

\begin{landscape}
  \begin{table}[h]
    \centering
    \caption{Goal 01 CPX: Implement an agile response strategy that allows the team to quickly pivot and adapt to emerging technologies and market demands. \codeIcon denotes metrics that were successfully automatically implemented.}
    \label{tab:g01cpx}
    \arrayrulecolor{grayline}
    \begin{minipage}[t]{0.67\textwidth}
      \small
      \subcaption{Metrics generated by \CMS}
      \label{tab:cms-metrics-g1cpx}
      \arrayrulecolor{grayline}
    \begin{tabularx}{\textwidth}{>{\hsize=0.2\hsize}X>{\hsize=0.8\hsize}X}
      \hfill\CMSIcon & Metrics generated by \CMS \\ \hline
    \codeIcon\texttt{Customer}-\texttt{Orientation}-\texttt{Index} & [\textellipsis] the extent of customer orientation of a team, calculated by aggregating and analyzing comments, reactions, pull requests, and issues data. Higher indices indicate greater attention to user-reported issues and faster response times. \\ \hline
    \texttt{Disruption}-\texttt{Adaptation}-\texttt{Index} & [\textellipsis] the frequency and speed at which a team adopts and integrates emerging technologies, calculated by the average time gap between consecutive adoptions. Faster adaptation indicates less time between adoptions. \\ \hline
    \codeIcon\texttt{Team}-\texttt{Learning}-\texttt{Index} & [\textellipsis] the growth in technical repertoire of the team over time, assessed by the variety and complexity of libraries and practices in repositories and their change over time. A higher index suggests a culture of continuous learning. \\ \hline
    \texttt{Calculate}-\texttt{ChangeAdaption}-\texttt{Rate} & [\textellipsis] the ratio of project elements adjusted and implemented due to emerging technologies or market demands to total project elements in a given time frame. \\ \hline
    \texttt{Calculate}-\texttt{Customer}-\texttt{Feedback}-\texttt{Implementation} & [\textellipsis] the ratio of customer feedback points implemented to the total number of feedback points received in a given time frame. \\ \hline
    \texttt{CalculateSprint}-\texttt{CompletionRate} & [\textellipsis] the ratio of completed tasks to total set tasks at the beginning of every sprint cycle, with a completed task defined as one checked off as completed in the sprint. \\ \hline
    \codeIcon\texttt{Adaptive}-\texttt{PlanningIndex} & [\textellipsis] the frequency of team check-ins, code pushes, and code reviews, plus the number of changes made in response to external tech trends, determined by commit or issue descriptions. \\ \hline
    \codeIcon\texttt{Integration}-\texttt{Delivery}-\texttt{Frequency} & [\textellipsis] how often new updates, features, or patches are integrated and delivered within the project, determined by commit history and pull request data. \\ \hline
    \codeIcon\texttt{Refactoring}-\texttt{Frequency} & [\textellipsis] the frequency of code refactoring, examined through commit history for keywords related to refactoring or significant code alterations. Commit comments and changes provide additional insights. \\ \hline
      \end{tabularx}
      \arrayrulecolor{black}
    \end{minipage}%
    \hspace{0.3em}
    \begin{minipage}[t]{0.58\textwidth}
      \small
      \subcaption{Metrics generated by \Vanilla GPT4API}
      \label{tab:vanilla-metrics-g1cpx}
      \arrayrulecolor{grayline}
    \begin{tabularx}{\textwidth}{>{\hsize=0.2\hsize}X>{\hsize=0.8\hsize}X}
        \hfill\vanillaIcon & Metrics generated by \Vanilla \\ \hline
        \texttt{branchUsage}-\texttt{Frequency} & [\textellipsis] measure how frequently the team uses branches. Agile teams often use branches for different features or tasks. \\ \hline
        \texttt{codeChurn} & [\textellipsis] measure the amount of code that is changed, added, or deleted over time. Agile teams should have low code churn rate as they aim for iterative and incremental development. \\ \hline
        \texttt{codeReview}-\texttt{Process} & [\textellipsis] measure the thoroughness of code reviews. Agile teams often emphasize on code reviews as part of their development process. \\ \hline
        \texttt{commit}-\texttt{Frequency} & [\textellipsis] measure the frequency of commits. Agile teams tend to commit changes often. \\ \hline
        \texttt{contributors}-\texttt{Count} & [\textellipsis] count the number of contributors. Agile teams usually have many contributors, indicating a high level of collaboration. \\ \hline
        \texttt{issue}-\texttt{Resolution}-\texttt{Time} & [\textellipsis] measure the average time taken to resolve issues. Agile teams should be able to resolve issues quickly. \\ \hline
        \texttt{pullRequest}-\texttt{Acceptance}-\texttt{Rate} & [\textellipsis] measure the acceptance rate of pull requests. High acceptance rate may indicate a well-coordinated team. \\ \hline
        \texttt{technology}-\texttt{Stack}-\texttt{Diversity} & [\textellipsis] measure the diversity of languages and technologies used in the repository. This could indicate the team's ability to adapt to different technologies. \\ \hline
        \texttt{testCoverage} & [\textellipsis] measure the extent of test coverage in the codebase. Agile teams usually emphasize on testing as part of their development process. \\ \hline
      \end{tabularx}
      \arrayrulecolor{black}
    \end{minipage}
  \end{table}
\end{landscape}

\begin{landscape}
  \begin{table}[h]
    \centering
    \caption{Goal 02 ABS: Foster an environment where every team member feels connected and informed. \codeIcon denotes metrics that were successfully automatically implemented.}
    \label{tab:g02abs}
    \arrayrulecolor{grayline}
    \begin{minipage}[t]{0.67\textwidth}
      \small
      \subcaption{Metrics generated by \CMS}
      \label{tab:g02abs-cms-metrics}
      \arrayrulecolor{grayline}
      \begin{tabularx}{\textwidth}{>{\hsize=0.2\hsize}X>{\hsize=0.8\hsize}X}
        \hfill\CMSIcon & Metrics generated by \CMS \\ \hline
        \texttt{Evaluate}-\texttt{Adaptability} & [\textellipsis] team's adaptability by determining how often and how effectively changes are incorporated into the projects. [\textellipsis] analyze the frequency and nature of commits and pull requests along with the corresponding comments and reactions. [\textellipsis] measure how quickly and smoothly new issues are created and resolved, and how new ideas, suggested in the form of pull requests or issues, are accepted and merged into the projects. \\ \hline
        \texttt{Measure}-\texttt{Communication}-\texttt{Flow} & [\textellipsis] level of interactions among team members by analyzing the frequency, variety, and response time of comments, reviews, and reactions in the repositories. [\textellipsis]  the number of comments and responses in pull requests, issues, and commit reviews, along with the number of positive reactions to these interactions. \\ \hline
        \codeIcon\texttt{Track}-\texttt{Collaborative}-\texttt{Efforts} & [\textellipsis] track the degree of collaborations of team members by calculating the average number of contributors per commit, per issue, and per pull request. [\textellipsis] assignments of tasks to gather info on how workload is distributed among team members. [\textellipsis] measure the activities of joint problem solving in issues and pull requests. \\ \hline
        \codeIcon\texttt{Calculate}-\texttt{Cultural}-\texttt{InclusivityScore} & [\textellipsis] calculated by analyzing the various types of GitHub interactions among team members captured in the data. This can include the number of comments made by each member, the diversity of projects they contribute to, and the receptivity of their contributions (measured through reactions, mentions, or replies). \\ \hline
        \codeIcon\texttt{Determine}-\texttt{CareerAnchor}-\texttt{Alignment} & [\textellipsis] the correlation between a team member's career anchors and their contributions on GitHub. Though not explicit in the database, career anchors can be inferred by looking at the tasks individuals choose, the complexity of tasks they handle, and their growth in area of contribution. [\textellipsis] \\ \hline
        \codeIcon\texttt{Measure}-\texttt{Continuous}-\texttt{LearningGrowth} & [\textellipsis] quantifies the learning curve of each team member by tracking how their contributions evolve over time in terms of complexity (measured by lines of code), diversity (measured by project involvement), and acceptance (measured by merge rate for pull requests). Additionally, the number of issues raised and resolved, and the growth in the breadth of their engagement (number of different repositories contributed to) can be used as indicators of learning. \\ \hline
        \codeIcon\texttt{Alignment}-\texttt{Score} & Measure the alignment of team members on project goals and tasks.[\textellipsis] such as pull requests, issues resolved, and consistency in commit messages. This score will be higher if there is consistency in the activity revolving around the project goals. \\ \hline
        \codeIcon\texttt{Communication}-\texttt{Frequency} & Measure the frequency of interactions between team members. [\textellipsis] include comments, reactions, and reviews on a particular project. [\textellipsis] obtained from tables such as those storing issue discussion or pull request discussion data. [\textellipsis] \\ \hline
        \texttt{DiversityScore} & Calculate a diversity score based on the variety of contributors in a project. [\textellipsis]  geographical location, affiliation, and role of each contributor, inferred by user metadata and association with organizations. Teams with diverse contributors would have a high diversity score. \\ \hline
      \end{tabularx}
      \arrayrulecolor{black}
    \end{minipage}%
    \hspace{0.3em}
    \begin{minipage}[t]{0.58\textwidth}
      \small
      \subcaption{Metrics generated by \Vanilla}
      \label{tab:g02abs-vanilla-metrics}
      \arrayrulecolor{grayline}
      \begin{tabularx}{\textwidth}{>{\hsize=0.2\hsize}X>{\hsize=0.8\hsize}X}
        \hfill\vanillaIcon & Metrics generated by \Vanilla \\ \hline
        
        \texttt{Active}-\texttt{Discussions} & Counts the number of active discussions on pull requests and issues. Active discussions can indicate good communication within the team. \\ \hline
        \texttt{CodeChurn} & Measures the amount of code that is rewritten or deleted after being written. Higher churn might indicate a lack of clarity or agreement within the team. \\ \hline
        \texttt{CodeReview}-\texttt{Frequency} & Estimates the frequency of code reviews. Regular code reviews indicate a team that shares knowledge and improves together. \\ \hline
        \texttt{CommitComment}-\texttt{Frequency} & Counts the number of comments made on commits. Comments can be a sign of active discussion and feedback.\\ \hline
        \texttt{Commit}-\texttt{Frequency} & Measures the frequency of commits. Regular commits indicate good collaboration and active engagement. \\ \hline
        \texttt{Contributors}-\texttt{Diversity} & Assesses the diversity of contributors in terms of the number of different people making commits, opening issues, and submitting pull requests. A diverse range of contributors may suggest a more inclusive environment. \\ \hline
        \texttt{Issue}-\texttt{Resolution}-\texttt{Time} & Measures the average time taken to resolve issues. Faster resolution times can suggest a well-coordinated team. \\ \hline
        \texttt{PullRequest}-\texttt{Frequency} & Assesses the frequency of pull requests. Frequent pull requests can indicate a collaborative environment. \\ \hline
        \texttt{Repository}-\texttt{StarGazers} & Counts the number of stars the repository has. This can be a sign of the team's impact within the organization. \\ \hline
      \end{tabularx}
      \arrayrulecolor{black}
    \end{minipage}
  \end{table}
\end{landscape}

\begin{landscape}
  \begin{table}[h]
    \centering
    \caption{Goal 02 CPX: Implement a structured, cross-platform communication system enabling real-time collaboration and knowledge sharing among all software development teams. \codeIcon denotes metrics that were successfully automatically implemented.}
    \label{tab:g02cpx}
    \arrayrulecolor{grayline}
    \begin{minipage}[t]{0.67\textwidth}
      \small
      \subcaption{Metrics generated by \CMS}
      \label{tab:g02cpx-cms-metrics}
      \arrayrulecolor{grayline}
      \begin{tabularx}{\textwidth}{>{\hsize=0.2\hsize}X>{\hsize=0.8\hsize}X}
        \hfill\CMSIcon & Metrics generated by \CMS \\ \hline
        \texttt{CalculateDevice}-\texttt{AndOsCompat}-\texttt{ibility} & This function calculates the ratio of successfully completed activities (like commits, pull requests, etc.) to attempted activities across different devices and operating systems. This could be extracted through logs capturing activity metadata. \\ \hline
        \texttt{CalculateInfor}-\texttt{mationFind}-\texttt{ability} & This function calculates the average time taken by a user to find a piece of information or perform an action in the system. This could potentially be obtained from system logs capturing timestamped user behaviours. \\ \hline
        \texttt{CalculateUser}-\texttt{Interactions} & This function calculates the count of interactions a user has with the platform. It includes all activities like comments, commits, pull requests, issues raised, reviews, reactions etc. The metric can be calculated by summing up all actions performed by a user in the platform across all tables storing user information. \\ \hline
        \texttt{AnalyzeObject}-\texttt{OrientedDesign}-\texttt{Adoption} & This method should analyze the commit history and associated files to identify if object-oriented principles are adopted. It can be achieved by detecting signature practices of object-oriented programming such as encapsulation, inheritance, and polymorphism in the code and seeing patterns of their application in the development process. \\ \hline
        \texttt{AssessCollab}-\texttt{orationAndKnow}-\texttt{ledgeSharing} & This function should be designed to examine the user interactions data i.e., comments, reactions, review data, and assignees associations, to evaluate the level of collaboration and knowledge sharing happening. Milestones can be analyzed for cooperative planning, PR and issues comments for knowledge sharing, and user reactions for feedback loops. The degree of interconnectedness between teams and within teams can also be quantified. \\ \hline
        \texttt{ComputeUnified}-\texttt{Modeling}-\texttt{Utilization} & This function should scrutinize the historical record of commits and associated metadata files to determine the extent of UML. This can involve searching for specific file formats (.uml, .puml, .xmi, etc.) and evaluating the frequency and recency of their updates. \\ \hline
        \codeIcon\texttt{AverageIssue}-\texttt{ResolutionTime} & This metric calculates the average time taken to resolve an issue. This can be calculated from the timestamp of when an issue is opened to when it is closed. \\ \hline
        \texttt{Continuous}-\texttt{Integration}-\texttt{Frequency} & This metric measures the frequency of integrations into the main branch done by the teams. This can be calculated by taking count of all the merged pull requests into the main branch over a defined period (monthly, quarterly, etc). \\ \hline
        \codeIcon\texttt{UserEngage}-\texttt{ment} & This metric measures the level of user engagement in the project. User engagement can be quantified in terms of the number of comments, reactions, reviews on issues and pull requests from each user over a defined timeframe. \\ \hline
    
      \end{tabularx}
      \arrayrulecolor{black}
    \end{minipage}%
    \hspace{0.3em}
    \begin{minipage}[t]{0.58\textwidth}
      \small
      \subcaption{Metrics generated by \Vanilla}
      \label{tab:g02cpx-vanilla-metrics}
      \arrayrulecolor{grayline}
      \begin{tabularx}{\textwidth}{>{\hsize=0.2\hsize}X>{\hsize=0.8\hsize}X}
        \hfill\vanillaIcon & Metrics generated by \Vanilla \\ \hline
    \texttt{Average}-\texttt{ResponseTime}-\texttt{ToIssues} & This function will calculate the average response time to the issues raised. It would help you understand how quickly the team is able to address the issues. \\ \hline
    \texttt{CodeChurn} & This function will calculate the rate at which code is added, modified, or deleted. It can help you understand the stability of the project. \\ \hline
    \texttt{FrequencyOf}-\texttt{Commits} & This function will calculate the frequency of commits in the repository. Frequent commits usually indicate an active development process and good collaboration among the team. \\ \hline
    \texttt{NumberOf}-\texttt{Branches} & This function will return the number of branches in the repository. Multiple branches often suggest that the team is capable of working on different features or tasks simultaneously. \\ \hline
    \texttt{NumberOf}-\texttt{Contributors} & This function will return the total number of contributors to the repository. It will help understand the team size and its potential capacity. \\ \hline
    \texttt{NumberOfForks} & This function will return the number of forks. It can give you an idea about the number of people or teams who are interested in their projects or using their code. \\ \hline
    \texttt{NumberOf}-\texttt{IssuesOpened}-\texttt{VsClosed} & This function will compare the number of opened issues versus closed ones. It will help you understand how effectively the team is solving the issues. \\ \hline
    \texttt{NumberOfPull}-\texttt{Requests} & This function will return the number of pull requests. This metric can be useful to understand the team's code review process and how often they merge their code. \\ \hline
    \texttt{NumberOfStars} & This function will return the number of stars. It can give you an idea about the popularity or the quality of the project. \\ \hline
    \texttt{NumberOf}-\texttt{Watchers} & This function will return the number of watchers. It can give you an idea about the number of people interested in updates to the project. \\ \hline
      \end{tabularx}
      \arrayrulecolor{black}
    \end{minipage}
  \end{table}
\end{landscape}

\begin{landscape}
  \begin{table}[h]
    \centering
    \caption{Goal 03 ABS: Ensure every piece of code released is tested thoroughly and confidently.s \codeIcon denotes metrics that were successfully automatically implemented.}
    \label{tab:g03abs}
    \arrayrulecolor{grayline}
    \begin{minipage}[t]{0.67\textwidth}
      \small
      \subcaption{Metrics generated by \CMS}
      \label{tab:g03abs-cms-metrics}
      \arrayrulecolor{grayline}
      \begin{tabularx}{\textwidth}{>{\hsize=0.2\hsize}X>{\hsize=0.8\hsize}X}
        \hfill\CMSIcon & Metrics generated by \CMS \\ \hline
        \texttt{EarlyFault}-\texttt{Detection}-\texttt{Rate} & This metric measures the percentage of total detected faults found during the early stages of the testing process. This could be calculated by analyzing the timestamps of fault detection within the test case commit histories and categorizing them into early, middle, and late stages of the testing process. \\ \hline
        \texttt{RegressionTest}-\texttt{Coverage} & This metric measures the percentage of code modifications covered by regression tests in each release. This requires tracking data on code modifications (which can be derived from commit histories) and data on the test cases selected for regression testing. \\ \hline
        \texttt{TestSuite}-\texttt{Reduction}-\texttt{Rate} & This metric measures the percentage reduction in the number of test cases used to test a release compared to the previous one. This would require tracking the number of tests run for each code release, which can be inferred from the interactions and commit history related to test cases in the repository. \\ \hline
        \codeIcon\texttt{Calculate}-\texttt{Feedback}-\texttt{Response}-\texttt{Time} & This function calculates the mean time taken to respond to feedback such as bug reports or code review comments. Inputs to the function could be an array of timestamps for when feedback was provided and an array of timestamps for when the feedback was addressed. \\ \hline
        \texttt{Calculate}-\texttt{Pipeline}-\texttt{Efficiency} & This function calculates the efficiency of the Continuous Integration/Continuous Deployment (CI/CD) pipeline by computing the mean time taken between each commit to being successfully built and tested in the CI/CD environment. Inputs to the function could be an array of timestamps for each commit and an array of timestamps for when the commit was built and tested. \\ \hline
        \texttt{MeasureTest}-\texttt{Coverage} & This function measures the percentage of codebase that is covered by automated tests. Inputs to the function could be the total number of lines in the codebase and the number of lines that have been tested using automated tests. \\ \hline
        \texttt{AutomatedTest}-\texttt{Coverage} & This metric will calculate the percentage of the code that is covered by automated tests. It can be computed by dividing the number of lines of code executed during automated testing by the total lines of code in the system. \\ \hline
        \codeIcon\texttt{Performance}-\texttt{Issue}-\texttt{Frequency} & This metric calculates the frequency of performance-related issues in the software system. It can be computed by dividing the total number of performance-related issues by the total number of issues opened in a given time period. \\ \hline
        \texttt{SelfHealing}-\texttt{Events}-\texttt{Count} & This metric calculates the number of instances where the software system has automatically detected and rectified faults. It requires tracking events where self-healing protocols are triggered in the system. \\ \hline
      \end{tabularx}
      \arrayrulecolor{black}
    \end{minipage}%
    \hspace{0.3em}
    \begin{minipage}[t]{0.58\textwidth}
      \small
      \subcaption{Metrics generated by \Vanilla}
      \label{tab:g03abs-vanilla-metrics}
      \arrayrulecolor{grayline}
      \begin{tabularx}{\textwidth}{>{\hsize=0.2\hsize}X>{\hsize=0.8\hsize}X}
        \hfill\vanillaIcon & Metrics generated by \Vanilla \\ \hline
        \texttt{CodeChurn} & Code churn is the amount of code that is rewritten or deleted soon after it's written. High code churn may indicate that the team is struggling with the implementation or there are a lot of bugs in the code. \\ \hline
        \texttt{CodeCoverage} & This is a measure of the amount of code that is covered by automated tests. High code coverage is a good sign that the team is testing thoroughly. \\ \hline
        \texttt{CodeQuality}-\texttt{Metrics} & Analyze the code for complexity, maintainability, duplications, and issues in the code. This will help to understand the quality of code that is being produced. \\ \hline
        \texttt{Commit}-\texttt{Frequency} & This shows the activity and pace of the team. More frequent commits mean the team is actively working and delivering. \\ \hline
        \texttt{IssueMetrics} & The number of open vs closed issues, the time taken to close them, and the issue resolution rate can provide insights into the team's efficiency and effectiveness. \\ \hline
        \texttt{PullRequest}-\texttt{Metrics} & Analyzing the number of pull requests, their acceptance rate, the time taken to merge them, and the number of comments on them can provide insight into the team's development and review process. \\ \hline
    
      \end{tabularx}
      \arrayrulecolor{black}
    \end{minipage}
  \end{table}
\end{landscape}

\begin{landscape}
  \begin{table}[h]
    \centering
    \caption{Goal 3 CPX: Establish a multi-layered testing framework that incorporates unit testing, integration testing, and automated end-to-end testing, ensuring maximum coverage and efficiency. \codeIcon denotes metrics that were successfully automatically implemented.}
    \label{tab:g03cpx}
    \arrayrulecolor{grayline}
    \begin{minipage}[t]{0.67\textwidth}
      \small
      \subcaption{Metrics generated by \CMS}
      \label{tab:g03cpx-cms-metrics-g5cpx}
      \arrayrulecolor{grayline}
    \begin{tabularx}{\textwidth}{>{\hsize=0.2\hsize}X>{\hsize=0.8\hsize}X}
      \hfill\CMSIcon & Metrics generated by \CMS \\ \hline
        \texttt{AutomatedTest}-\texttt{Ratio} & This calculates the ratio of automated tests to total tests. The data required would be the number of automated test cases and total test cases. Both of these can be inferred from commit records and issue data. \\ \hline
        \codeIcon\texttt{Defect}-\texttt{Density} & To compute the defect density, count how many issues have been raised (implying a defect was found) divided by the total size of the codebase (which can be approximated by the number of commits or lines of code). \\ \hline
        \texttt{TestCoverage}-\texttt{Ratio} & This metric should calculate the ratio of the number of unit tests to the number of functions or methods in the code. The data required would be the number of unit test cases (which can be inferred from the commits mentioning addition or modification of test cases) and the number of functions or methods (which can be inferred from the code within the commits). \\ \hline
        \texttt{AutomatedEndTo}-\texttt{EndTest}-\texttt{Frequency} & This metric measures how frequently automated end-to-end tests are carried out. It can be calculated by monitoring the testing suite, and recording the timestamps of each automated end-to-end test. \\ \hline
        \texttt{IntegrationTest}-\texttt{Failures} & This metric captures the number of integration test failures over a given period. This could be calculated by examining the logs of the testing suite, identifying instances of failures in integration tests. \\ \hline
        \texttt{UnitTest}-\texttt{Coverage} & This metric measures the percentage of code that is covered by unit tests. This can be calculated by identifying the sections of code that are executed during unit tests, and dividing by the total lines of code. \\ \hline
        \texttt{CalcCI-CD}-\texttt{Integration}-\texttt{Frequency} & This function measures the frequency of integration in the CI/CD pipeline. It can be determined by the commit history and the occurrence of automatic builds and tests. Each cycle in the CI/CD pipeline triggered by a commit can be considered an integration instance. \\ \hline
        \texttt{CalcTest}-\texttt{Coverage} & This function will calculate the test coverage ratio, defined as the number of lines of code executed during testing divided by the total number of lines in the codebase. This can be inferred by the commit history and the areas of the codebase that changes are being made to. Any change that involves the test suite can be counted towards the lines of code tested. \\ \hline
        \texttt{CountDocumented}-\texttt{Tests} & This function counts the number of test cases that have corresponding documentation, inferred from the commit history. Any commit that includes changes or additions to the documentation and is linked to a specific test can be counted. \\ \hline
    
      \end{tabularx}
      \arrayrulecolor{black}
    \end{minipage}%
    \hspace{0.3em}
    \begin{minipage}[t]{0.58\textwidth}
      \small
      \subcaption{Metrics generated by \Vanilla GPT4API}
      \label{tab:g03cpx-vanilla-metrics}
      \arrayrulecolor{grayline}
    \begin{tabularx}{\textwidth}{>{\hsize=0.2\hsize}X>{\hsize=0.8\hsize}X}
        \hfill\vanillaIcon & Metrics generated by \Vanilla \\ \hline
        \texttt{CodeQuality} & Tools like CodeClimate can provide a maintainability index for the repository which can indicate the quality of code written by the team. \\ \hline
        \texttt{CommitActivity} & This refers to the frequency of updates to the repository. Regular commits indicate a team's active involvement and consistent work. \\ \hline
        \texttt{Contributors} & The number of contributors can give an idea of the size of the team and how many people are actively working on the project. \\ \hline
        \texttt{Issues} & The number of issues opened and closed can give an indication of the team's ability to identify and resolve problems. \\ \hline
        \texttt{PullRequests} & The number of pull requests created and merged can be an indicator of the team's collaboration level and their ability to accomplish tasks. \\ \hline
        \texttt{TestCoverage} & If the team has integrated any code coverage tools like Codecov, Coveralls with their Github repository, you can check the percentage of the code base covered by their tests. This would directly indicate how seriously they take testing. \\ \hline
      \end{tabularx}
      \arrayrulecolor{black}
    \end{minipage}
  \end{table}
\end{landscape}

\begin{landscape}
  \begin{table}[h]
    \centering
    \caption{Goal 04 ABS: Deliver high-quality products on time, every time, without burning out our team. \codeIcon denotes metrics that were successfully automatically implemented.}
    \label{tab:g04abs}
    \arrayrulecolor{grayline}
    \begin{minipage}[t]{0.67\textwidth}
      \small
      \subcaption{Metrics generated by \CMS}
      \label{tab:g04abs-cms-metrics}
      \arrayrulecolor{grayline}
      \begin{tabularx}{\textwidth}{>{\hsize=0.2\hsize}X>{\hsize=0.8\hsize}X}
        \hfill\CMSIcon & Metrics generated by \CMS \\ \hline
        \codeIcon\texttt{Agile}-\texttt{Adherence} & This function calculates the team's adherence to agile practices by measuring the frequency of changes made (number of commits), the responsiveness to issues (time taken to close an issue after it is opened), and the degree of customer collaboration (number of interactions with users or clients on issues and pull requests). \\ \hline
        \texttt{Communication}-\texttt{Frequency} & This function calculates the frequency and diversity of communication between team members. It measures this by counting the total number of interactions among team members (comments on commits, issues, and pull requests) and the number of unique individuals involved in these interactions in a given period. \\ \hline
        \texttt{TeamWellbeing} & This function assesses team well-being by looking at indicators like frequency of late-night commits (as potential signs of overtime or burnout) and the distribution of work among team members (variability in the number of commits made by each team member). \\ \hline
        \codeIcon\texttt{Delivery}-\texttt{Time}-\texttt{Efficiency} & This metric should represent the average time taken for issues or pull requests marked with specific labels (indicating the type and/or complexity of the task) to close. A lower average time may indicate efficient delivery. This metric should control for the complexity as inferred by labels associated with issues or pull requests. \\ \hline
        \texttt{FeedbackLoop}-\texttt{Efficiency} & This metric should quantify the average response time to issues or pull requests. A shorter average response time would indicate that feedback is provided more quickly, enabling a rapid feedback loop. This metric should consider each interaction as a possible feedback mechanism - comments, reviews, reactions, etc. \\ \hline
        \codeIcon\texttt{TeamComposit}-\texttt{ionBalance} & This metric should evaluate the balance of skills within a team. This can be indirectly inferred from the different types of contributions (commits, reviews, comments, issues created/resolved) by different users in a repository. A more diverse distribution of contribution types may indicate a balance of skills within the team. \\ \hline
        \codeIcon\texttt{MeasureCode}-\texttt{Refactor}-\texttt{Frequency} & This metric measures the frequency of code refactors by each team member over a given period of time. It is a count of the number of refactoring-related commits made per developer. \\ \hline
        \texttt{MeasureCommit}-\texttt{Frequency} & This metric measures the frequency of git commits by each team member over a given time period. It is a count of the number of commits made per developer. \\ \hline
        \codeIcon\texttt{Measure}-\texttt{Issue}-\texttt{ResolutionTime} & This metric calculates the average time taken to resolve issues. It involves timestamp values when issues were opened and the moment they were closed. The resolution time of an issue is the difference between these two timestamps. The average resolution time can be computed over all resolved issues. \\ \hline
      \end{tabularx}
      \arrayrulecolor{black}
    \end{minipage}%
    \hspace{0.3em}
    \begin{minipage}[t]{0.58\textwidth}
      \small
      \subcaption{Metrics generated by \Vanilla}
      \label{tab:g04abs-vanilla-metrics}
      \arrayrulecolor{grayline}
      \begin{tabularx}{\textwidth}{>{\hsize=0.2\hsize}X>{\hsize=0.8\hsize}X}
        \hfill\vanillaIcon & Metrics generated by \Vanilla \\ \hline
        \texttt{Burnout}-\texttt{Indicator} & To identify signs of team burnout. This could be measured by sudden decreases in commit frequency, increased time to resolve issues, or increased code churn. \\ \hline
        \texttt{CodeChurn} & To calculate the amount of code that gets modified over time. Lower churn might mean stable and mature code. \\ \hline
        \texttt{CodeCoverage} & To assess the percentage of code covered by tests. Higher coverage could imply a lower likelihood of bugs and better quality. \\ \hline
        \texttt{CodeReviews} & To measure the frequency and thoroughness of code reviews. Regular, detailed reviews can improve code quality and reduce bugs. \\ \hline
        \texttt{Commit}-\texttt{Frequency} & To measure the frequency and consistency of code contribution. Regular commits may indicate a healthy and active team. \\ \hline
        \texttt{IssuesResolut}-\texttt{ionTime} & To gauge how long it takes for the team to resolve issues. Shorter times could suggest efficiency and good problem-solving skills. \\ \hline
        \texttt{OpenToClosed}-\texttt{IssueRatio} & To understand the balance between new issues and resolved issues. A lower ratio may indicate a team that effectively handles issues. \\ \hline
        \texttt{PullRequest}-\texttt{Acceptance}-\texttt{Rate} & To evaluate the percentage of pull requests that are accepted. A higher rate may reflect a team that produces high-quality code. \\ \hline
        \texttt{Release}-\texttt{Frequency} & To track how often new versions or updates are released. More frequent releases might suggest a team that can deliver products on time. \\ \hline
      \end{tabularx}
      \arrayrulecolor{black}
    \end{minipage}
  \end{table}
\end{landscape}

\begin{landscape}
  \begin{table}[h]
    \centering
    \caption{Goal 04 CPX: Adopt an Agile project management approach, facilitating flexible planning, progressive development, early deployment, and continuous improvement. \codeIcon denotes metrics that were successfully automatically implemented.}
    \label{tab:g04cpx}
    \arrayrulecolor{grayline}
    \begin{minipage}[t]{0.67\textwidth}
      \small
      \subcaption{Metrics generated by \CMS}
      \label{tab:cms-metrics-g04cpx}
      \arrayrulecolor{grayline}
    \begin{tabularx}{\textwidth}{>{\hsize=0.2\hsize}X>{\hsize=0.8\hsize}X}
      \hfill\CMSIcon & Metrics generated by \CMS \\ \hline
        \texttt{AdaptiveIterat}-\texttt{ionScore} & [\textellipsis] adapting to changes and improvements in their agile methodology. [\textellipsis] by tracking the changes in the number, type, or magnitude of features committed in each iteration (which can be represented by commit records within a certain timeframe). More variation in feature commitments might suggest a more adaptive approach. \\ \hline
        \texttt{Cooperative}-\texttt{Communication}-\texttt{Score} & [\textellipsis] extent of communication and collaboration among all stakeholders. [\textellipsis] summing up the total number of user interactions from commit comments, issue discussions, and pull request reviews, then dividing by number of unique users. [\textellipsis] pulled from the comment, issue, and review tables, respectively. [\textellipsis] \\ \hline
        \texttt{UserRequirement}-\texttt{FulfilmentRatio} & [\textellipsis] quantifies how well the development efforts are aligning with user requirements. [\textellipsis] comparing the number of closed issues linked with a user story or a use case against the total number of issues created. Issues can be identified in the issue table with status and link to user story (if available), exact metric calculation may vary based on data available. \\ \hline
        \texttt{Calculate}-\texttt{UrgencyIndex} & [\textellipsis] calculate an urgency index by evaluating the frequency and intensity of discussions related to project development and enhancements in project comments, issue discussions, and pull request reviews. [\textellipsis] gauged through sentiment analysis and keyword usage. \\ \hline
        \texttt{Determine}-\texttt{ClarityOfVision} & [\textellipsis] measure the clarity of vision within the team by analyzing the consistency of language used in project descriptions, PR descriptions, and issue descriptions for project development with Agile principals. Natural language processing can be applied to provide a metric for this purpose. \\ \hline
        \texttt{Measure}-\texttt{Empowerment}-\texttt{Level} & [\textellipsis] assess the empowerment level of the team members by examining the distribution of contributions(stats of code contributions, PRs, reviews made etc.) and decision making(with open issue discussions and closed PRs) amongst the team members. [\textellipsis] calculated as a distribution metric for individual contributors. \\ \hline
        \texttt{Calculate}-\texttt{Iterative}-\texttt{Development} & [\textellipsis] average time taken for each development iteration, which can be inferred by the time difference between consecutive commits by the same user. The average should be calculated across all users over a specified period of time. \\ \hline
        \texttt{Measure}-\texttt{Communication}-\texttt{Effectiveness} & [\textellipsis] effectiveness of communication in a team by calculating the average number of comments per issue opened within a specific period. [\textellipsis] \\ \hline
        \texttt{Track}-\texttt{Continuous}-\texttt{Improvement} & [\textellipsis] the ratio of closed issues to the total number of issues opened within a specified period. A higher ratio indicates that the team is effectively addressing and resolving issues, signifying continuous improvement. \\ \hline

      \end{tabularx}
      \arrayrulecolor{black}
    \end{minipage}%
    \hspace{0.3em}
    \begin{minipage}[t]{0.58\textwidth}
      \small
      \subcaption{Metrics generated by \Vanilla GPT4API}
      \label{tab:vanilla-metrics-g04cpx}
      \arrayrulecolor{grayline}
    \begin{tabularx}{\textwidth}{>{\hsize=0.2\hsize}X>{\hsize=0.8\hsize}X}
        \hfill\vanillaIcon & Metrics generated by \Vanilla \\ \hline
        \texttt{Branching}-\texttt{Strategy} & This function analyzes the team's branching strategy. In Agile, developers often work on separate branches and merge their changes frequently, promoting flexible planning and early deployment. \\ \hline
        \texttt{CodeReview}-\texttt{Process} & This function assesses how often code is reviewed and who is involved in the code review process. Regular code reviews are an important part of Agile methodologies, promoting continuous improvement and high-quality code. \\ \hline
        \texttt{Commit}-\texttt{Frequency} & This function provides information on how often the team commits to the repository. Agile teams often make smaller, more frequent commits, which allows for more flexible planning and continuous improvement. \\ \hline
        \texttt{Contributor}-\texttt{Diversity} & This function measures the diversity of contributors to the repository. Agile teams often have a wide range of contributors, reflecting their emphasis on teamwork and collaboration. \\ \hline
        \texttt{Issue}-\texttt{Resolution}-\texttt{Time} & This function calculates the average time taken to close issues. Agile teams should be able to respond quickly to issues, reflecting their capacity for flexible planning. \\ \hline
        \texttt{PullRequest}-\texttt{Volume} & This function measures the number of pull requests made by the team. Frequent pull requests are indicative of early deployment and progressive development. \\ \hline
        \texttt{TestCoverage} & This function provides information about the percentage of the codebase covered by automated tests. High test coverage is usually a sign of an Agile team, as this allows for continuous improvement and reduces the risk of regressions. \\ \hline
      \end{tabularx}
      \arrayrulecolor{black}
    \end{minipage}
  \end{table}
\end{landscape}

\end{document}